\newcounter{myctr}
\def\myitem{\refstepcounter{myctr}\bibfont\noindent\ifnum\themyctr>9\else\phantom{0}\fi\hangindent17pt\themyctr.\enskip}

\documentclass{ws-ijqi}
\usepackage{hyperref}
\usepackage[super,sort,compress]{cite}

\newcommand{\ket}[1]{\ensuremath{\left|#1\right\rangle}}
\newcommand{\bracket}[2]{\ensuremath{\left\langle#1 \vphantom{#2}\right| \left. #2 \vphantom{#1}\right\rangle}}

\begin{document}

\catchline{}{}{}{}{}

\title{On Wavefunction Collapse, \\ 
the Einstein-Poldolsky-Rosen Paradox \\ 
and Measurement in Quantum Mechanics and Field Theory}

\author{Stuart Samuel\footnote{Retired Professor 
\newline
\hspace*{1.5 mm}Previous Institutions (in historical order):  
\newline
\hspace*{5 mm} Institute for Advanced Study at Princeton 
\newline
\hspace*{5 mm} Columbia University
\newline
\hspace*{5 mm} City College of New York
\newline
E-mail: stuartsamuel@hotmail.com}}

\maketitle


\begin{abstract}

We first consider the Einstein-Podolsky-Rosen (EPR) paradox 
for the system of two particles with spin ${1 \over 2}$ with entangled spins
in first-quantized quantum mechanics. 
If measurement is governed by wavefunction collapse, 
then we are able to show using gedanken experiments
that a number of fundamental principles 
including conservation of angular momentum and the Heisenberg uncertainty principle can be violated. 
We conclude that the collapse of the spin part of the wavefunction cannot happen 
and therefore an EPR paradox does not arise for this system. 
Indeed, quantum mechanical unitarity alone is sufficient to rule out ``spooky'' action at a distance.
The absence of spin wavefunction collapse 
leads to several interesting conclusions 
about how measurement works in quantum mechanics: 
When wavefunction collapse does not happen, 
(i) a signal from a macroscopic measuring devices indicating that a system is in a state $s$ 
does not necessarily mean that it is or was in $s$ 
and 
(ii) the uncertainty in quantum mechanics at the microscopic level is transmitted 
to uncertainty in signals for the macroscopic measuring device. 
In addition, we derive two general results on how quantum measurement works
and illustrate them in an example measuring the spin of a spin-${1 \over 2}$ object. 
In contrast to first-quantized quantum mechanics, 
in quantum field theory we are unable to fully rule out the possibility of wavefunction collapse. 
However, when we insist on linearity and unitarity -- thereby excluding the possibility of wavefunction collapse --
and additionally take into account in the wavefunction
all degrees of freedom involved in the measurement (including those of the equipment, experimentalists, etc.), 
a consistant acceptable quantum-mechanical understanding 
of the Bell-inequality-type experiments involving entangled photon polarizations emerges 
including the absence of an EPR paradox. 
It is interesting that the ability of an experimentalist to be {\it aware} of his (or her) perceptions 
but not be able to be conscious of other components in a linear superposition of states 
of a generalized Schr\"odinger Cat plays a role 
in resolving the Measurement Problem in quantum mechanics and
in making unitary quantum field theory consistent with observation.

\vspace{5mm} 
\end{abstract}


\markboth{Stuart Samuel}
{Wavefunction Collapse, the EPR Paradox and QM Measurement}


\section{Introduction to the EPR Paradox}
\label{intro}

The Einstein-Podolsky-Rosen\cite{epr} (EPR) paradox is a quantum mechanical effect that has bothered many physicists. 
In the original 1935 paper, Einstein, Podolsky and Rosen argued that quantum mechanics is either incomplete or inconsistent. 
The authors consider two particles that interact for a while and then move far apart but in such a way as to have their positions correlated. 
More precisely, they consider a wavefunction $\Psi({\bf x}_1, {\bf x}_2)$ such that if the first particle is located at ${\bf x}_1$ 
then at later times the position ${\bf x}_2$ of the second particle is ${\bf x}_2 = {\bf x}_1 + {\bf x}_0$ where ${\bf x}_0$ is a large distance, 
so large that the two particles no longer significantly interact. 
Such a situation might occur if a particle at rest decays into two particles. 
One observer then measures precisely the position of the first particle to be some value 
${\bf x}_1^f$ while a second observer measures precisely the momentum ${\bf p}_2^f$ of the second particle. 
From the first measurement, one deduces the position ${\bf x}_2^f$ of the second particle to be 
 ${\bf x}_2^f = {\bf x}_1^f + {\bf x}_0$ thereby simultaneously determining the momentum ${\bf p}_2^f$ and position ${\bf x}_2^f$ 
of the second particle at some later final time in violation of the Heisenberg uncertainty principle. 

The standard argument for avoiding the violation of the Heisenberg uncertainty principle uses wavefunction collapse. 
A measurement on the first particle not only causes its wavefunction to collapse
but also causes the wavefunction of the second particle to do so 
thereby destroying the correlations between the two particles. 
However, this bothered Einstein, Podolsky and Rosen: 
How can the measurement performed on the first particle cause a wavefunction change for the second faraway particle? 
This would seem to violate causality in special-relativity. 
However, there is a theorem\cite{NoCommunicationA,NoCommunicationB,NoCommunicationC,NoCommunicationD} that states that 
no message or communication between the two observers can be sent using this effect. 

\begin{figure}[h]
\centerline{\includegraphics[width=11.242cm, height=4.888cm]{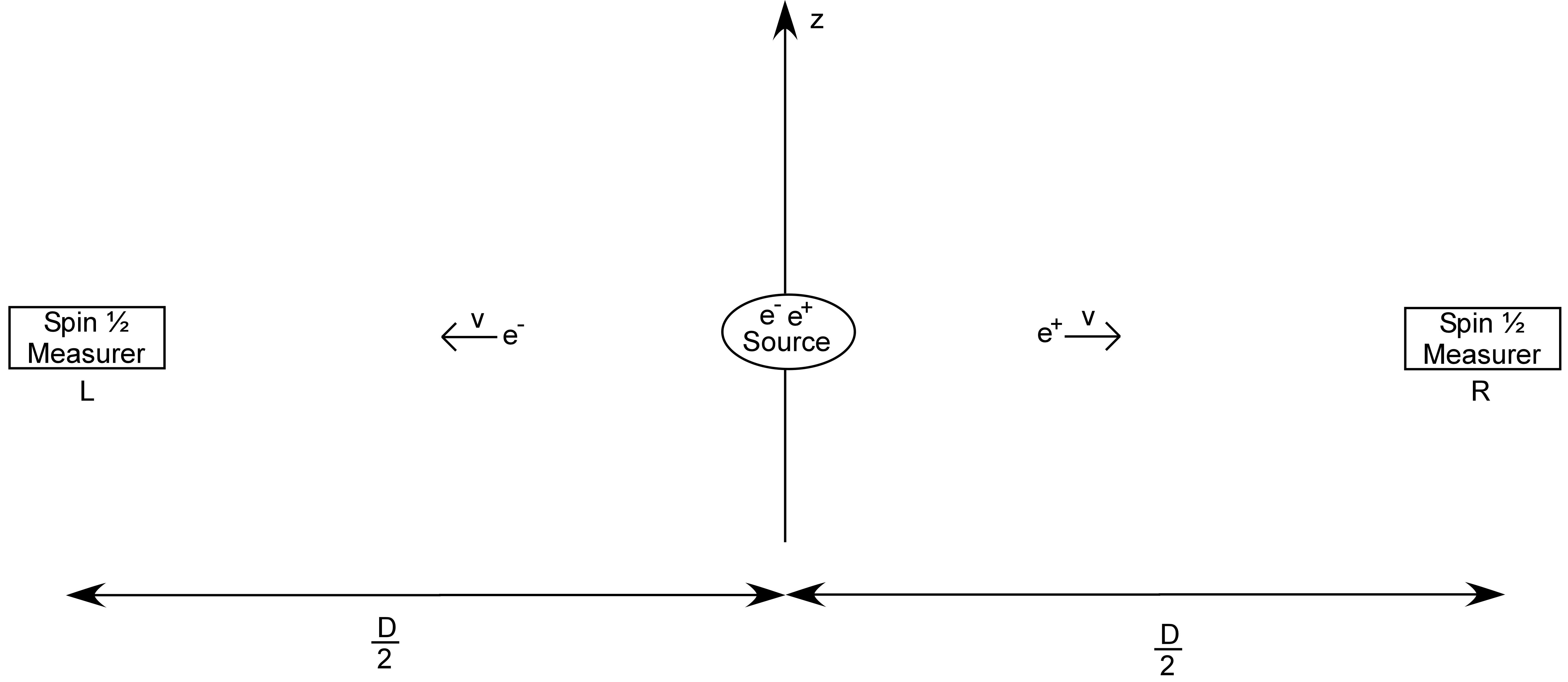}}
\vspace*{8pt}
\caption{Setup for the entangled spin-$1 \over 2$ EPR paradox.}
\label{fig:EPRsetup}
\end{figure}

An EPR-like paradox can be produced for any system of two objects involving two non-commuting observables, 
in which the observables are correlated over a long distance. 
The quintessential theoretical system involves the entanglement of two spin-${1 \over 2}$ particles:\cite{EntangledSpin} 
Two particles are generated at a central location with one moving to the left and one moving to the right, 
and with a combined spin angular momentum of $0$: 
\begin{equation}
\ket{00} = ( \ket{\uparrow}_{-} \ket{\downarrow}_{+} -  \ket{\downarrow}_{-} \ket{\uparrow}_{+})/\sqrt{2}
\, .
\label{SpinZeroState}
\end{equation}
Here, the subscript $-$ (respectively $+$) denotes the particle moving to the left or in the negative direction 
(respectively right or the positive direction), $\ket{00}$ indicates the state with $S = S_z  = 0$, and $\ket{\uparrow}$ 
denotes a spin-$1 \over 2$ state with $S_z = +{1 \over 2}$, that is, $ \ket{ \frac{1}{2}, \frac{1}{2} }$,  
and $\ket{\downarrow}$ denotes a state with $S_z = -{1 \over 2}$, that is, $\ket{ \frac{1}{2},-\frac{1}{2} }$. 
Here we used standard notation:\cite{quantummechanicsBookDirac,quantummechanicsBookBorn} 
$\ket{S S_z}$ represents an eigenstate with total spin $S$ and 
a value of $S_z$ for the $z$-component of spin. 
The entangled spin state in Eq.(\ref{SpinZeroState}) can arise for any two spin-${1 \over 2}$ particles 
but for the purposes of clarity and presentation,  
we choose the particle moving to the left to be an electron 
and the particle moving to the right to be a positron. 
In this case, the $-$ and $+$ can also stand for electron and positron. 
The setup is shown in Figure \ref{fig:EPRsetup}. 

The EPR paradox for this entangled spin-${1 \over 2}$ system is as follows: 
The electron and positron are allowed to speed away from one another until they are very far apart. 
Then a measurement is made on the electron to determine its spin in the $z$-direction. 
Just before the measurement, the spin of the positron could be up or down.
If, after the measurement, the spin of the electron is up then the spin of the positron must be down;
and if the electron's spin is down then the positron's spin must be up. 
There are two disturbing aspects about this. 
(i) First, from the measurement of the electron's spin on the left, one immediately knows the spin of the positron on the right. 
If the time it takes to measure the electron's spin is $\Delta t_M$ 
and the distance $D$ between the two particles is greater than $\Delta t_M /c$, 
then knowledge about the positron's spin is obtained faster than it takes light to propagate from the electron to the positron. 
(ii) Second, the measurement of the electron's spin on the left 
appears to have forced the spin of the positron to take on a particular value. 
It seems that a measurement on the electron is causing an effect 
on the faraway positron's (spin) wavefunction that, again, 
can happen faster than the time it takes light to propagate between the two particles.  
This second issue (``spooky action at a distance'') is often considered as the main EPR paradox.\footnote{When we refer to the EPR paradox 
without specifying which aspect, we mean issue (ii).}

Aspect (i) of the EPR paradox involves the propagation of {\it knowledge} 
(but not information\cite{NoCommunicationA,NoCommunicationB,NoCommunicationC,NoCommunicationD}) faster than the speed of light. 
It is straightforward to show that there is nothing paradoxical about this. 

\begin{figure}[h]
\centerline{\includegraphics[width=9.25cm, height=6.59cm]{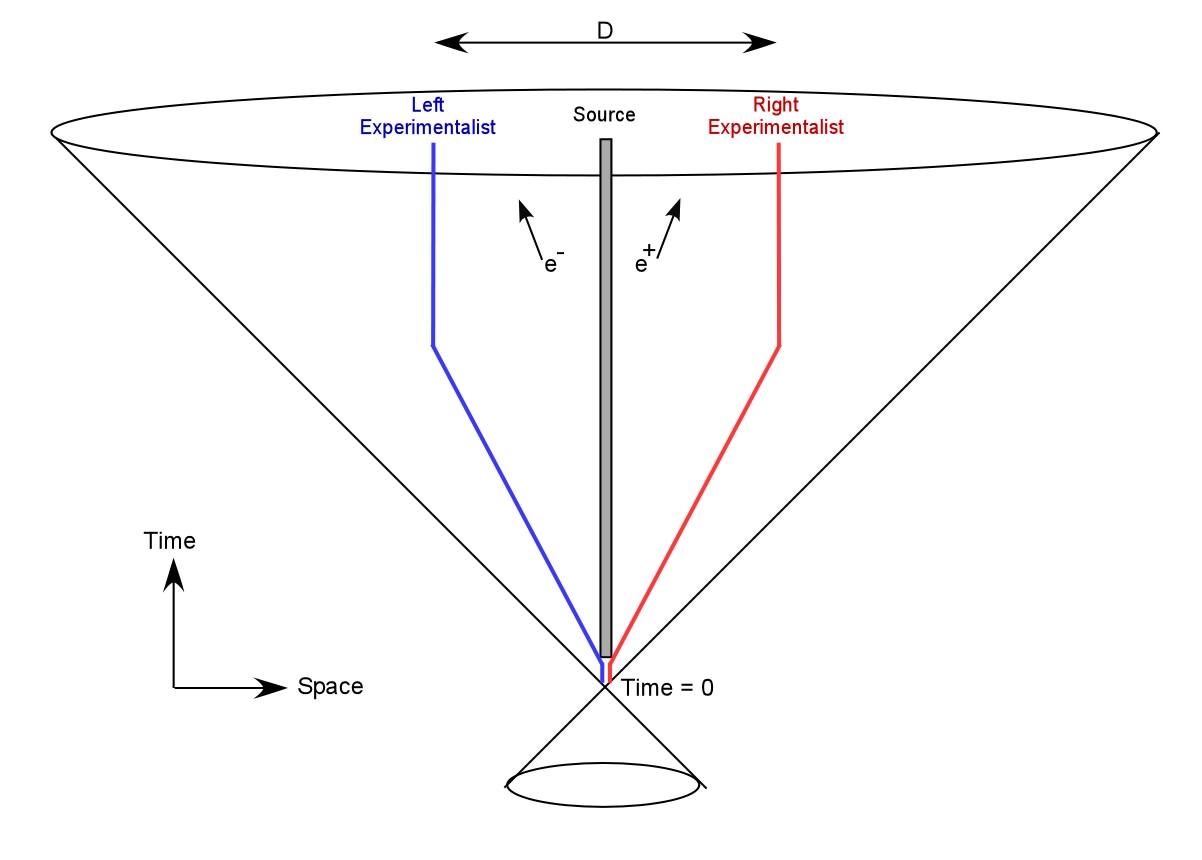}}
\vspace*{8pt}
\caption{The space-time light cone containing 
the events of the entangled spin-$1 \over 2$ experiment.}
\label{fig:LightCone}
\end{figure}

Essential to the instantaneous knowledge of the spin of the positron 
by the observer at the left is a prior agreement about the experimental setup 
and that the entire experiment lies within the causal future light cone of this agreement event. 
Figure \ref{fig:LightCone} shows the situation. 
In this figure, the two experimentalists get together at $t = 0$, 
the source is positioned at the origin and arranged to generate anti-correlated spins for the electron and positron, 
one experimentalist moves to the left and setups up a spin detector there, 
while the other experimentalist does the same to the right, 
the positron and electron are emitted at some later time from the source, and finally their spins are measured. 
Since everything happens within the future causal light cone, 
the rapid transfer of knowledge of the spin of the position does not violate special relativity. 
To emphasize the point, consider  the situation in Figure \ref{fig:LightCone} in which the experimentalists 
do not discuss the nature of the setup and do not know that the source will generate correlated spins. 
When the experimentalist at the left measures the electron's spin, 
the experimentalist then does not know that the positron's spin is opposite to the electron's spin 
even though the source for the electron and positron is creating anti-correlated spins; 
no instantaneous knowledge of the far-away-spin is gleaned in this case. 
Hence, aspect (i) is non-problematic. 
The analysis in this paragraph applies to other versions of the EPR paradox 
including the one in the original EPR paper: 
EPR setups can convey knowledge about distant objects faster than the speed of light 
but this is not in violation of special relativity. 

The spin-${1 \over 2}$ measuring devices in Figure \ref{fig:EPRsetup} are assumed to be able to measure the spin in the $z$-direction. 
Let us assume that the spin measurement involves spin wavefunction collapse: 
Suppose that a particle is in a linear combination of $s_z$ spin states, that is, 
the spin part of the wavefunction is $a\ket{\uparrow} + b \ket{\downarrow}$. 
Then, if the device measures the $z$-component of a spin-${1 \over 2}$ particle to be up, 
then at some point in the measurement process the state becomes  $a' \ket{\uparrow}$
and this is achieved by destroying the down component of $z$-spin $b  \ket{\downarrow} \rightarrow 0$ 
and renormalizing the $a' \ket{\uparrow}$ wavefunction: $\bracket {a'}{a'} = 1$.
Similar statements hold when the device measures the spin to be down. 
We also assume that the spin measuring device is ``accurate'' meaning that 
the probability of measuring up spin is $\bracket{a}{a} / N$ and the probability of measuring down spin is $\bracket{b}{b} / N$.
Here, $N =\bracket{a}{a} + \bracket{b}{b}$, and $\bracket{c}{c}$ indicates the inner product of the wavefunction factor $c$ 
for all degrees of freedom except for the particle's spin component. 
The objects $a$, $b$ and $c$ involve the spatial wavefunction of the particle 
as well as the spin states and wavefunctions of all other entities relevant to the experiment 
including those of the measuring device and the nearby environment. 
Wavefunction collapse is one interpretation of measurement in quantum mechanics.\cite{quantummechanicsBookDirac,WHeisenberg,JvonNeumann}
We refer to a measurement of spin of the nature described in this paragraph 
as a \textit{Spin-${1 \over 2}$ Collapse Measurement}. 

The Spin-$1 \over 2$ Collapse Measurement in Figure \ref{fig:EPRsetup} creates aspect (ii) of the EPR paradox: 
the measurement of the spin of the electron leads to a collapse of 
${( \ket{\uparrow}_- \ket{\downarrow}_+ -  \ket{\downarrow}_- \ket{\uparrow}_+ ) / \sqrt{2}}$
to either $\ket{\uparrow}_- \ket{\downarrow}_+$  
or to $\ket{\downarrow}_- \ket{\uparrow}_+$ each with a probability of 50\% 
thereby affecting the spin wavefunction of the faraway positron. 
Before the measurement on the electron, 
the positron had a $50\%$ chance of pointing up and a $50\%$ chance of pointing down. 
After the Spin-$1 \over 2$ Collapse Measurement, 
the spin of the positron is $100\%$ up if the spin of the electron has been measured to be down 
and $100\%$ down if the spin of the electron has been measured to be up. 
Hence, a long-distance change in the nature of the spin of the positron 
has been caused by the Collapse Measurement of the electron's spin.

Collapse of the wavefunction violates quantum-mechanics unitarity. 
The justification\cite{JvonNeumann} for this during a measurement 
is based on the connection between the wavefunction and probability: 
If a single measurement definitively determines that the spin is up (respectively, down) then 
the wavefunction must collapse and be proportional to $\ket{\uparrow}$ (respectively,  $\ket{\downarrow}$ ). 
 \textbf{The Converse Collapse Statement}, namely, \textit{if wavefunction collapse does not occur in nature 
then a single measurement cannot necessarily definitively determine the spin state}, follows from pure logic 
(if $A$ then $ B \Rightarrow$ If (not $B$) then (not $A$)). 
In such a case, multiple measurements may be needed to determine a quantum state. 
A further analysis, which is given in the body of this work, then determines the criterion 
as to when a single measurement suffices:
If wavefunction collapse is not needed ``to explain'' an experimental result, 
then a single measuring event suffices,
but if a measurement needs wavefunction collapse to explain it then multiple measurements 
are required to determine the quantum state.
We call this principle the {\bf Quantum Measurement Rule}.
It also turns out that 
multiple measurements are often needed to uncover an underlying quantum state 
even when wavefunction collapse does occur in nature; 
the criterion is the same as above but the reason is different. 
We illustrate the Quantum Measurement Rule for spin ${1 \over 2}$ in Section \ref{SGmeasurement}.

The first part of our work focuses on first-quantized quantum mechanics as opposed to second-quantized quantum mechanics (or quantum field theory),
the main difference being that the latter allows for the production and destruction of particles. 
In Section \ref{SecondQuantization}, we present our analysis for quantum field theory.

\section{Possible Violation of Angular Momentum in a Spin-${1 \over 2}$ Collapse Measurement of Entangled Spins}
\label{AngularMomentumViolation}

If a measurement creates a collapse of the wavefunction as described in Section \ref{intro}, 
then conservation of angular momentum can be violated as we now show: 
Assume that the initial state is an eigenstate of angular momentum with values of $J$ and $J_z$
as measured at the spatial point where the pair are created. 
Note that the initial $e^-$--$e^+$ pair and everything else in the universe 
are required to combine to give an angular momentum state of $\ket{J J_z}$.\footnote{The is a Gedanken experiment 
since this condition is not satisfied in real experimental situations.
} 
The initial eigenfunction for the system is 
$\Psi(\ket{\uparrow}_-\ket{\downarrow}_+ - \ket{\downarrow}_-\ket{\uparrow}_+)/\sqrt{2}$ 
where $\Psi$ is the wavefunction for the degrees of freedom of the electron, the positron and 
everything else except the spins of the electron and the positron. 
The angular momentum state of $\Psi$ must be $\ket{J J_z}$
since $(\ket{\uparrow}_-\ket{\downarrow}_+ - \ket{\downarrow}_-\ket{\uparrow}_+)$ is a $J = 0$ object. 
When the Spin-${1 \over 2}$ Collapse Measurement of the electron is performed, 
the wavefuntion collapses to either $\Psi'\ket{\uparrow}_-\ket{\downarrow}_+$ (if the electron's spin is measured to be up) 
or $\Psi''\ket{\downarrow}_-\ket{\uparrow}_+$ (if the electron's spin is measured to be down). 
Neither of these wavefunctions can be an eigenstate of angular momentum. 

The problem with either $\Psi'\ket{\uparrow}_-\ket{\downarrow}_+$ or $\Psi''\ket{\downarrow}_-\ket{\uparrow}_+$ 
as final states is that  $\ket{\uparrow}_-\ket{\downarrow}_+$ and $\ket{\downarrow}_-\ket{\uparrow}_+$ 
are a combination of spin $0$ and spin $1$, 
and it is not possible to construct a state with the original values of $J$ and $J_z$. 
To create a state with angular momentum $\ket{J J_z}$ for the spin $1$ part, 
either $\ket{\uparrow}_-\ket{\uparrow}_+$ or $\ket{\downarrow}_-\ket{\downarrow}_+$ 
or both must be present. 
For example, in the first case above, 
$\Psi'\ket{\uparrow}_-\ket{\downarrow}_+ = \Psi'(\ket{\uparrow}_-\ket{\downarrow}_+ - 
\ket{\downarrow}_-\ket{\uparrow}_+)/2 + \Psi'(\ket{\uparrow}_-\ket{\downarrow}_+ + \ket{\downarrow}_-\ket{\uparrow}_+)/2$. 
For the first term to have angular momentum content of $\ket{J J_z}$, $\Psi'$ must have angular momentum content of $\ket{J J_z}$ 
because $(\ket{\uparrow}_-\ket{\downarrow}_+ - \ket{\downarrow}_-\ket{\uparrow}_+)/2$ is spin $0$. 
For the second term to have angular momentum content of $\ket{J J_z}$, 
there must be present terms of the form $\Psi'_-\ket{\uparrow}_-\ket{\uparrow}_+$ and/or
$\Psi'_+\ket{\downarrow}_-\ket{\downarrow}_+$ 
where $\Psi'_-$ and $\Psi'_+$ have $J_z$ values of respectively $-1$ and $1$ and total angular momentum
of either $J-1$, $J$ or $J+1$; 
then angular momentum  $J-1$, $J$ or $J+1$ can in principle combine with angular momentum $1$ to give angular momentum $J$. 
However, since $\Psi'$ has to have total angular momentum $J$ because of the spin zero term, 
only the case of $J$ (and not $J-1$ or $J+1$) can occur for  $\Psi'_-$ and $\Psi'_+$.
For this to produce an angular momentum eigenstate $\ket{J J_z}$, 
$\Psi'$, $\Psi'_-$ and $\Psi'_+$ need to be related in the appropriate way with normalizations related to Clebsch-Gordon coefficients. 
Since $\Psi'_-$ and $\Psi'_+$ are absent if the spin part of the wavefunction collapses to $\Psi'\ket{\uparrow}_-\ket{\downarrow}_+$, 
angular momentum is violated. 
The second case in which the collapse is to $\Psi''\ket{\downarrow}_-\ket{\uparrow}_+$ also violates angular momentum.

Methods of measuring the spin of the electron must distinguish up spin from down spin.
Hence, the process of measuring the spin is likely to lead to a wavefunction of the form
$\Psi'\ket{\uparrow}_-\ket{\downarrow}_+ - \Psi''\ket{\downarrow}_-\ket{\uparrow}_+$ 
where $\Psi'$ differs from $\Psi''$.
In such cases, ``spin-flipping'' terms proportional to $\ket{\uparrow}_-\ket{\uparrow}_+$ and/or $\ket{\downarrow}_-\ket{\downarrow}_+$ 
have to arise when the initial state is an eigenstate of angular momentum;  
the reasoning is similar as in the previous paragraph. 
So not only cannot the wavefunction collapse to $\Psi'\ket{\uparrow}_-\ket{\downarrow}_+$ 
but ``spin-flipping'' terms proportional to $\ket{\uparrow}_-\ket{\uparrow}_+$ and/or $\ket{\downarrow}_-\ket{\downarrow}_+$ 
have to arise if $\Psi' \ne \Psi''$.

The violation of angular momentum can be verified by direct calculation. 
For $\Psi'\ket{\uparrow}_-\ket{\downarrow}_+$ to have a $z$-component of angular momentum of $J_z$,
it suffices for $\Psi'$ to have a $z$-component of angular momentum of $J_z$. 
The problem is not with $J_z$ but with total angular momentum. 
When $J^2$ is applied to $\Psi'\ket{\uparrow}_-\ket{\downarrow}_+$, 
it needs to produce $j(j+1) \Psi'\ket{\uparrow}_-\ket{\downarrow}_+$.
In the standard way, it is useful to  express $J^2$ as $J^2_z + (J_+ J_- +  J_- J_+)/2$ 
where $J_+ = J_z + i J_y$ is the $J_z$-raising operator and $J_- = J_z - i J_y$ is the $J_z$-lowering operator. 
Substitute for $J_i$ in $J^2$, $J_i = J'_i + S^{e^-}_i + S^{e^+}_i$ where $i = x, y$, $z$, $-$ or $+$, 
and where the $S^{e^-}_i$ (respectively, $S^{e^+}_i$) are the spin operators of the electron (respectively, positron).
Here, $J_i'$ is the $i^{th}$ component of angular momentum 
involving all the other degrees of freedom including those of the environment and 
those for the orbital angular momentum of the electron and positron. 
When this substitution is performed, 
many terms proportional to $\ket{\uparrow}_-\ket{\downarrow}_+$ are produced. 
One problematic term comes from 
$S^{e^-}_- S^{e^+}_+ \Psi'\ket{\uparrow}_-\ket{\downarrow}_+ = 
\hbar^2  \Psi'\ket{\downarrow}_-\ket{\uparrow}_+$ which needs to vanish and does not.
Other problematic terms come from 
$(J'_+S^{e^-}_- + J'_- S^{e^+}_+) \Psi'\ket{\uparrow}_-\ket{\downarrow}_+ = 
\hbar (J'_+  \Psi') \ket{\downarrow}_-\ket{\downarrow}_+ + \hbar (J'_- \Psi') \ket{\uparrow}_-\ket{\uparrow}_+$;
The first term $(J'_+  \Psi') \ket{\downarrow}_-\ket{\downarrow}_+$  is not zero unless $J'_z$ is a maximal value, 
that is, $J'_z = J'$,  
and the second term $( J'_- \Psi') \ket{\uparrow}_-\ket{\uparrow}_+$ is not zero unless $J'_z$ is a minimal value, 
that is,  $J'_z = -J'$.

If the initial state is an eigenstate with $J=0$ then the above argument does not work for a technical reason:\footnote{This was pointed out 
to the author by Stephen Adler.}
To measure spin in the $z$-direction, an axis in that direction must be picked out. 
Hence, it is not possible to have an initial state with $J=0$ and a measuring device for up and down spin.
This problem does not necessarily arise for $J>0$ and Appendix A presents an example.

\section{Problems with Casuality and the Uncertainty Principle in a Spin-${1 \over 2}$ Collapse Measurement of Entangled Spins}
\label{UncertaintyPrincipleViolation}
If the Spin-${1 \over 2}$ Collapse Measurement is first performed on the spin of the electron at the left 
in Figure \ref{fig:EPRsetup} and it is up (respectively down), 
then the spin part of the wavefunction collapses to $\ket{\uparrow}_-\ket{\downarrow}_+$ (respectively, $\ket{\downarrow}_-\ket{\uparrow}_+)$. 
Hence, when one subsequently measures the spin of the positron at the right, its spin will be down (respectively up). 
Similarly, statements hold for when the spin of the positron is first measured 
and the spin of the electron is measured afterwards. 
The perfect anti-spin correlation is borne out by Spin-${1 \over 2}$ Collapse Measurements. 
The same is true if one chooses to measure both spins in the $y$-direction because 
the form of the spin factor $(\ket{\uparrow}_-\ket{\downarrow}_+ - \ket{\downarrow}_-\ket{\uparrow}_+)$ 
is independent of the axis of spin quantization. 

If one first measures the spin of the electron in the $z$-direction and subsequently the spin of the positron in the $y$-direction, 
then the following occurs (assuming spin wavefunction collapse). 
In the case in which the electron spin measurement is up, then the wavefunction collapses to $\ket{\uparrow}_-\ket{\downarrow}_+$. 
The spin of the positron is down when using the $z$-axis for spin quantization, 
which is ${ (\ket{\uparrow}_+ - i\ket{\downarrow}_+)_y/\sqrt{2} }$ when using the $y$-axis for quantization, 
where the subscript indicates the axis of spin quantization. 
Hence, there will be a 50\% chance of subsequently measuring the $y$-spin of the positron as up 
and 50\% chance of measuring its $y$-spin as down. 
Furthermore, if the spin of the positron is up in the $y$-direction, 
it does not mean that one knows that the electron's spin is down in the $y$-direction 
because the collapse of the wavefunction has ruined the anti-correlation of spins in the $y$-direction: 
although 
$(\ket{\uparrow}_-\ket{\downarrow}_+ - \ket{\downarrow}_-\ket{\uparrow}_+)_z = 
  (\ket{\uparrow}_-\ket{\downarrow}_+ - \ket{\downarrow}_-\ket{\uparrow}_+)_y$ 
before the Spin-${1 \over 2}$ Collapse Measurement of the electron's spin, 
$(\ket{\uparrow}_-\ket{\downarrow}_+)_z = 
   (\ket{\uparrow}_- + i \ket{\downarrow}_-)_y(\ket{\uparrow}_+ - i\ket{\downarrow}_+)_y/2$ afterwards. 

In the above two paragraphs, 
the collapse of the wavefunction when measuring the electron's spin causes 
an instantaneous long-distance effect on the positron's spin state 
that prevents both (i)  inconsistencies in the measurement on the anti-alignment of the electron's and positron's spins 
and (ii) the simultaneous measurement of the $y$ and $z$ components of the electron's spin. 
These  two paragraphs are presented to contrast what happens in the rest of the paragraphs in this section. 

\begin{figure}[h]
\centerline{\includegraphics[width=1.0\textwidth]{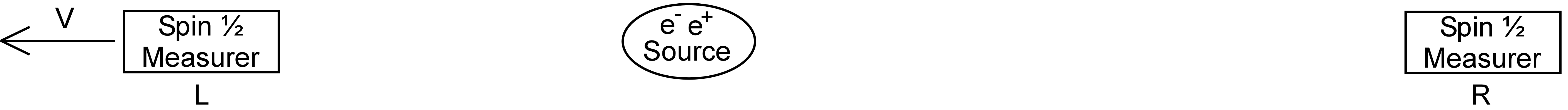}}
\caption{The entangled spin-${1 \over 2}$ setup with the left spin measurer (measuring device) 
moving at a very high speed $v$ to the left in the negative $x$-direction.}
\label{fig:setupMoving}
\end{figure}

\begin{figure}[h]
\centerline{\includegraphics[width=1.0\textwidth]{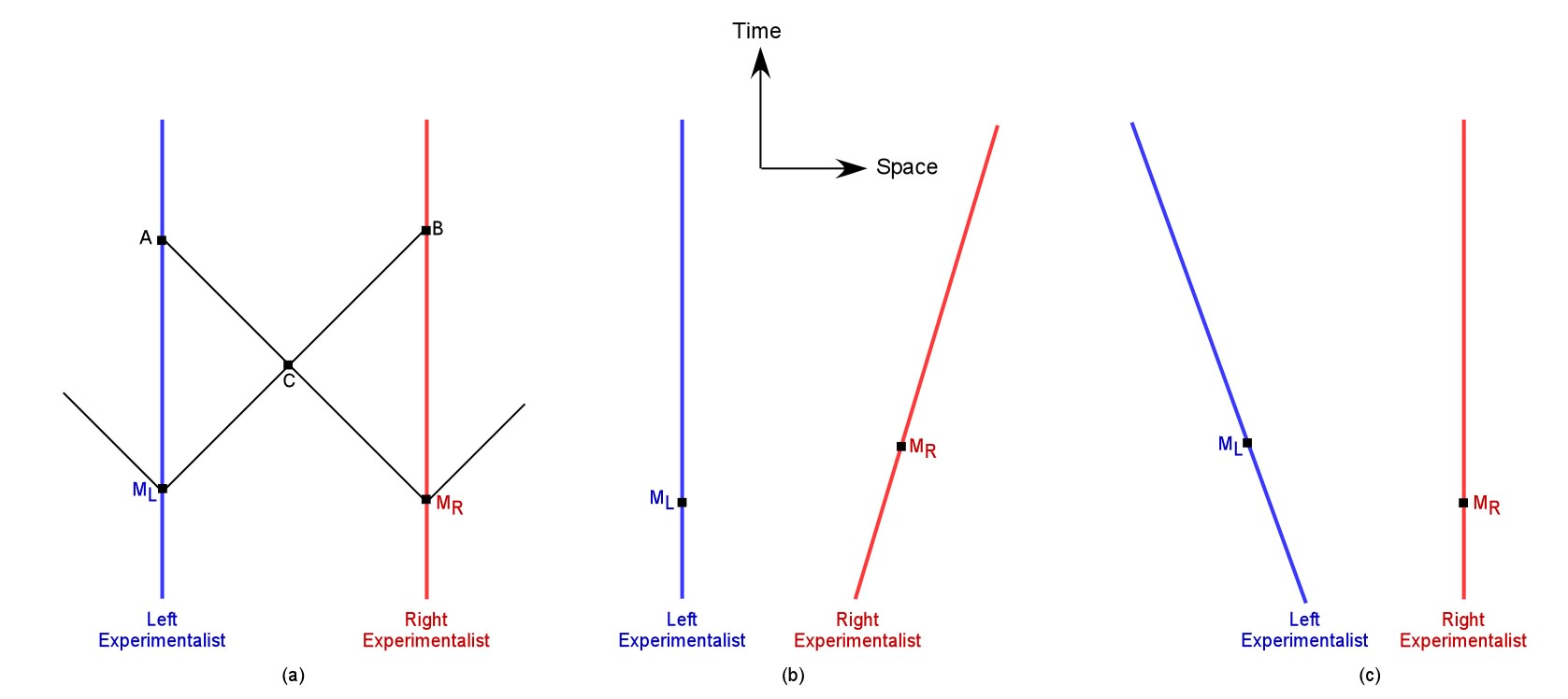}}
\caption{Different Reference Frames. The situation in Figure \ref{fig:EPRsetup}, 
for which the source and spin measuring devices have a common rest frame, corresponds to (a). 
For the case of the left spin measuring device moving relativistically fast as in Figure \ref{fig:setupMoving}, 
(b) corresponds to the rest frame of the left experimentalist
while (c) corresponds to the rest frame of the right experimentalist. 
In these space-time diagrams, $M_L$ (respectively $M_R$) is the space-time point at which the left (respectively, right) measurement takes place. 
Note that since it is impossible for the times associated with $M_L$ and $M_R$ to be exactly the same in (a),
the left measurement must occur before the right measurement or vice versa;  
then the standard argument is that wavefunction collapse (if it can happen) takes place when the first of these two measurements is made. 
This is problematic when devices are moving relativistically as (b) and (c) illustrate. 
The points $A$, $B$ and $C$ are used in the quantum field theory discussion in Section \ref{SecondQuantization}. 
The line from $M_L$ to $B$ is the path light would take if it were to be emitted at $M_L$ and aimed at the right-measuring device. 
For static measuring devices, the light arrives at $B$ in a space-time diagram. 
The line from $M_R$ to $A$ is similarly defined.}
\label{fig:differentReferenceFrames}
\end{figure}

Return to the setup in Figure \ref{fig:EPRsetup} but consider the case in which the experimentalist on the left and his measuring device 
are moving at a relativistic speed $v$ to the left. 
Also arrange things so that the experimentalist on the left is closer to the source 
when the electron-positron pair is emitted. 
We choose the velocity \textbf{v} to be in the (negative) $x$-direction, 
which eliminates the need to use Lorentz transformations in comparing spin measurements in the analysis below. 
The situation is shown in Figure \ref{fig:setupMoving}. 
Then due to special relativity, the speeds and initial positions can be adjusted so that each experimentalist claims 
to have made the spin measurement first. 
See Figures \ref{fig:differentReferenceFrames}(b) and \ref{fig:differentReferenceFrames}(c). 
For example, in the rest frame of Figure \ref{fig:setupMoving} (that is, the rest frame of the right observer and the source), 
when $v = 0.95 c$ and the speed of the electron and positron is $0.99 c$ and 
the distance of the left observer is $0.5 sec \ c$ (half a light-second) from the source and the right observer is $1.0 sec\  c$ from the source, 
then the measurement of the spin of the electron by the left observer in the right observer's rest frame happens 
almost $11.5$ seconds after the right observer has measured the spin of the positron (See Figure \ref{fig:differentReferenceFrames}(c)), 
while in the left observer's rest frame 
the measurement of the spin of the positron by the right observer happens almost $3$ seconds 
after he has measured the spin of the electron (See Figure \ref{fig:differentReferenceFrames}(b)). 
Henceforth, for presentation purposes, we take the experimentalist on the left to be male and the experimentalist on the right to be female.

When the experimentalist on the right measures the spin of the positron in the $z$-direction, 
the spin collapses to 
$(\ket{\uparrow}_-\ket{\downarrow}_+ - \ket{\downarrow}_-\ket{\uparrow}_+)/\sqrt{2} \rightarrow \ket{\downarrow}_-\ket{\uparrow}_+$ 
if she measures the spin of the positron to be up and collapses to 
$\ket{\uparrow}_-\ket{\downarrow}_+$ if  she measures it to be down. 
Both of these occur with $50\%$ probability and the outcome cannot be affected by the spin measurement of the electron because, in her rest frame, 
the spin measurement of the electron has not yet been made. 
See Figure \ref{fig:differentReferenceFrames}(c).
A similar situation exists for the experimentalist measuring the spin of the electron. 
In his rest frame, he sees the source and the other experimentalist moving rapidly away from him 
and the measurement of the spin of the positron has not yet taken place when he makes his measurement.
See Figure \ref{fig:differentReferenceFrames}(b).
So his measurement causes a collapse of 
$(\ket{\uparrow}_-\ket{\downarrow}_+ - \ket{\downarrow}_-\ket{\uparrow}_+)/\sqrt{2}  \rightarrow \ket{\uparrow}_-\ket{\downarrow}_+$ 
if he measures the electron spin to be up and to $\ket{\downarrow}_-\ket{\uparrow}_+$ 
if he measures it to be down. 
Each outcome occurs with $50\%$ probability and cannot be affected by the measurement of the positron on the right. 
Thus, there are four possible cases: two are compatible with each other with both measurements 
causing a collapse to $\ket{\uparrow}_-\ket{\downarrow}_+$ or 
to $\ket{\downarrow}_-\ket{\uparrow}_+$ and two are incompatible 
resulting in collapses by the right and left measurements that do not agree with each other. 
For example, both experimentalists can measure the spins to be up, 
meaning that the left experimentalist causes a collapse to $\ket{\uparrow}_-\ket{\downarrow}_+$ 
but the right experimentalist causes a collapse to $\ket{\downarrow}_-\ket{\uparrow}_+$. 
Thus, by using measuring devices that are moving with respect to each other, it is possible to create a causality contradiction. 
This does not happen for the non-moving case in Figure \ref{fig:EPRsetup}:
If the left measurement occurs first and wavefunction collapse is instantaneous and leads to $\ket{\uparrow}_-\ket{\downarrow}_+$
(respectively,  $\ket{\downarrow}_-\ket{\uparrow}_+$)
then the right experimentalist will measure the spin to be down (respectively, up). 
The argument is similar if the right measurement occurs first. 

Issues similar to the above have been raised in the context of entangled photons. See refs.[\refcite{suarez,scarani}]. 

When one has the left observer moving relativistically as in Figure \ref{fig:setupMoving}
and the left and right experimentalists measure spin in two perpendicular directions, 
it is possible to violate the Heisenberg uncertainty principle for a spin-${1 \over 2}$ state: 
\begin{equation}
		\Delta S_y\Delta S_z \ge {\hbar  \over 2} | \left\langle S_x \right\rangle | 
\ ,
\label{HeisenbergUncertaintyForSpinEq}
\end{equation}
as well as $\Delta S_z\Delta S_x \ge  \hbar  | \left\langle S_y \right\rangle | / 2$ and 
$\Delta S_x\Delta S_y \ge  \hbar  | \left\langle S_z \right\rangle | / 2$. 
These three conditions implie that at most only one of the spin components $S_x$, $S_y$ and $S_z$ can be measured with certainty.
Let the left experimentalist measure the $z$-component of the electron's spin 
and the right experimentalist measure the $y$-component of the positron's spin. 
Recall that the structure $(\ket{\uparrow}_-\ket{\downarrow}_+ - \ket{\downarrow}_-\ket{\uparrow}_+)$ 
is the same regardless of the direction of the axis of spin quantization. 
So if the right experimentalist uses a Spin-${1 \over 2}$ Collapse Measurement in Figure \ref{fig:setupMoving} and 
observes the $y$-component of spin of the positron to be up (respectively down) 
then the $y$-component of the spin of the electron must be down (respectively up). 
The right observer subsequently reports her experimental result to the left experimentalist. 
Since each experimentalist makes his or her measurement before the other one 
when the spin wavefunction is still $(\ket{\uparrow}_-\ket{\downarrow}_+ - \ket{\downarrow}_-\ket{\uparrow}_+)$, 
both can claim that their measurements are exact, valid and unaffected by the measurement of the other. 
In this way, the left experimentalist knows the $z$-component of the electron's spin from his measurement of it and 
deduces its $y$-component (since it is opposite to the result reported by the right experimentalist for the positron). 
Hence, both $y$- and $z$-components of the spin of the electron 
are determined precisely, which violates the Heisenberg uncertainty relation in Eq.(\ref{HeisenbergUncertaintyForSpinEq}). 
We label this as aspect (iii) of the EPR paradox. 
Aspect (iii) was also part of the original EPR paper\cite{epr} 
but the uncertainty relation involved position and momentum.

Note that the use of experimentalists moving with respect to one another is necessary 
to create the issues with wavefunction-collapse compatibility and with the Heisenberg uncertainty principle. 
In Figure \ref{fig:EPRsetup} in which the observers are not moving with respect to one another, 
the measurement of the left experimentalist happens before the measurement of the right experimentalist or vice versa 
in the rest frame of the experiment. 
See Figure \ref{fig:differentReferenceFrames}(a).
Similarly, in this same setup of Figure \ref{fig:EPRsetup}, 
as is evident from the first two paragraphs in this section, 
the  Heisenberg uncertainty principle is not violated because the wavefunction collapse of the first measurement 
necessarily affects the measurement of the second. 
In Figure \ref{fig:setupMoving}, this does not happen: 
although the collapse is instantaneous and non-local, 
both measurements of the experimentalists occur before the other in their respective rest frames 
as is illustrated in Figures \ref{fig:differentReferenceFrames}(b) and \ref{fig:differentReferenceFrames}(c).

\section{Absence of an EPR Paradox in First-Quantized Quantum Mechanics}
\label{EPRresolution}

Given that a Spin-$1 \over 2$ Collapse Measurement not only violates unitarity and linearity
but can also lead to a causal compatibility problem, 
and violate both conservation of angular momentum and the Heisenberg Uncertainty Principle for an object of spin $1 \over 2$,
it is reasonable to conclude that collapse of the spin part of the wavefunction 
cannot happen for the entangled spin-$1 \over 2$ system in first-quantized quantum mechanics. 
If collapse does not happen then the measurement of the spin of the electron does not force 
${( \ket{\uparrow}_- \ket{\downarrow}_+ -  \ket{\downarrow}_- \ket{\uparrow}_+ ) / \sqrt{2}}$ 
to become either $\ket{\uparrow}_- \ket{\downarrow}_+$  or 
 $\ket{\downarrow}_- \ket{\uparrow}_+$ and 
there is no long-distance action on the spin of the positron at the right 
from the measurement by the experimentalist on the left. 
This resolves aspect (ii) of the EPR paradox. 
Without wavefunction collapse, there is no ``spooky'' action at a distance. 

Although wavefunction collapse is able to avoid violation of the Heisenberg Uncertainty Principle 
for the ``static'' case (see the second paragraph of the previous section), 
the opposite is true for the situation using detectors moving relative to one another in Figure \ref{fig:setupMoving}.
Indeed, wavefunction collapse is the basis for the violation of the Heisenberg Uncertainty Principle in this case. 
This is not only true of the entangled spin-$1 \over 2$ system 
but the original situation considered by Einstein, Podolsky and Rosen\cite{epr} 
if relativistically moving measuring devices are used.

If collapse of the spin part of the wavefunction does not happen 
then the Converse Collapse Statement holds, namely that
it is not necessarily possible to determine the spin of an object in a single measurement. 
If a single measurement cannot definitively determine the spin then there is no violation of Eq.(\ref{HeisenbergUncertaintyForSpinEq}) 
and aspect (iii) of the EPR paradox is also resolved. 
We illustrate the need for multiple measurements to determine the spin of a state 
in Section \ref{SGmeasurement} below. 

Unitarity by itself is sufficient to show that aspect (ii) of the EPR paradox cannot arise 
for the entangled spin-${1 \over 2}$ system. 
As is well known, the evolution of a quantum system is generated by the operator 
\begin{equation}
U(t_f, t_0) = T[ Exp (- {i \over \hbar} \int_{t_0}^{t_f} H(s) ds ) ]
\, ,
\label{UnitaryOperator}
\end{equation}
which is the time-ordered exponential product of the Hamiltonian $H$. 
Here, $T [ ... ]$ indicates time ordering. 
When applied to the wavefunction at an initial time $t_o$, $U(t_f, t_o)$ produces the wavefunction at a later time $t_f$. 
As a unitary operator, $U(t_f, t_o)$ preserves the norms and orthogonality properties of states. 
In particular, if $\Psi_1(t_o)$ and $\Psi_2(t_o)$ are two states at time $t_o$ 
then $\Psi_1(t_f)  = U(t_f, t_o)\Psi_1(t_o)$ and $\Psi_2(t_f) = U(t_f, t_o)\Psi_2(t_o)$
are the solutions to the Schr\"{o}dinger equation (namely, $i \hbar \partial \Psi(t)/\partial t = H(t)\Psi(t)$) at time $t_f$.
These wavefunctions satisfy 
\begin{equation}
\begin{aligned}
\vert \Psi_1(t_f) \vert^2 & = \vert \Psi_1(t_o) \vert^2   \ \ ,   \\
\vert \Psi_2(t_f) \vert^2 & = \vert \Psi_2(t_o) \vert^2   \ \ ,   \\
 {\langle  \Psi_1(t_f) \vert   \Psi_2(t_f) \rangle } & =  {\langle  \Psi_1(t_o) \vert   \Psi_2(t_o) \rangle }
\, .
\label{Unitarity}
\end{aligned}
\end{equation}
In other words, quantum mechanical evolution preserves the norms and inner products of wavefunctions. 

To see whether a measurement of the electron at the left is able to affect the spin of the positron at the right, 
have the positron travel very far away in Figure \ref{fig:EPRsetup}, 
insist that there be no long-distance interactions between the electron and the positron,  
and require the positron to be in an insolated region 
where nothing nearby affects its spin. 
We shall show that if the measurement of the electron respects unitarity then the spin of the positron is unaffected. 
The initial wavefunction is 
$\Psi {( \ket{\uparrow}_- \ket{\downarrow}_+ -  \ket{\downarrow}_- \ket{\uparrow}_+ )}$
where $ {\langle  \Psi \vert   \Psi \rangle } = {1 \over 2}$. 
The probability that the positron's spin is up is 50\% and the probability that its spin is down is 50\%. 
This is not only true for when the axis of quantization for spin is in the $z$-direction 
but also true when this axis is in any direction. 
This characterizes the nature of the spin of the positron since any measurement on it 
is only able to produce these ``50\% -- 50\%'' results.
We need to show that this property remains true after a unitary-respecting measurement on the electron is made. 

Because nothing in the region nearby the positron can affect its spin, unitarity takes the following form: 
$\Psi \ket{\uparrow}_- \ket{\downarrow}_+ \to \Psi_f  \ket{\downarrow}_+$ and 
$\Psi \ket{\downarrow}_- \ket{\uparrow}_+ \to \Psi'_f  \ket{\uparrow}_+$ where 
$\Psi_f$ and $\Psi'_f$ are wavefunctions at any time after the measurement on the electron has been performed; 
these wavefunctions include everything except the spin of the positron and satisfy 
 $ {\langle  \Psi_f \vert   \Psi_f \rangle } =  {\langle  \Psi'_f \vert   \Psi'_f \rangle } =  {\langle  \Psi \vert   \Psi \rangle } = {1 \over 2}$ 
and ${\langle  \Psi_f \vert   \Psi'_f \rangle } = 0$  if unitarity holds.
Therefore, $\Psi {( \ket{\uparrow}_- \ket{\downarrow}_+ -  \ket{\downarrow}_- \ket{\uparrow}_+ )} \to 
({ \Psi_f  \ket{\downarrow}_+ -  \Psi'_f  \ket{\uparrow}_+ })$.
The probability of having the positron's spin up is still 50\% and the probability of having it down is still 50\%, 
a result that only depends on 
$ {\langle  \Psi_f \vert   \Psi_f \rangle } =  {\langle  \Psi'_f \vert   \Psi'_f \rangle } = {1 \over 2}$.

If the quantization axis for spin of the positron is taken to be in a direction $\bf{n}$ that is different from the $z$-axis, 
then expressing $\ket{\downarrow}_+ = \alpha \ket{\downarrow}_{\bf{n}} + \beta \ket{\uparrow}_{\bf{n}}$ 
and  $\ket{\uparrow}_+ = \beta^* \ket{\downarrow}_{\bf{n}} - \alpha^* \ket{\uparrow}_{\bf{n}}$ 
for some complex numbers satisfying $\alpha^*\alpha + \beta^*\beta = 1$, 
one finds $\Psi {( \ket{\uparrow}_- \ket{\downarrow}_+ -  \ket{\downarrow}_- \ket{\uparrow}_+ )} \to
( \alpha \Psi_f - \beta^*\Psi'_f ) \ket{\downarrow}_{\bf{n}} + (\beta \Psi_f + \alpha^*\Psi'_f ) \ket{\uparrow}_{\bf{n}}$.
A short calculation reveals that the probability of having the positron's spin up in the direction $\bf{n}$ 
is still 50\% and the probability of having it down is still 50\%; 
this result not only depends on $ {\langle  \Psi_f \vert   \Psi_f \rangle } =  {\langle  \Psi'_f \vert   \Psi'_f \rangle } = {1 \over 2}$ 
but {\it also} ${\langle  \Psi_f \vert   \Psi'_f \rangle } = 0$. 
Hence, the nature of the positron's spin is unchanged.
In this paragraph, $\ket{\uparrow}_{\bf{n}}$ and $\ket{\downarrow}_{\bf{n}}$ 
denote respectively up and down spin when the axis of quantization for spin is in the $\bf{n}$ direction.

This completes the proof that unitarity, which is a fundamental property of first-quantized quantum mechanics, 
is sufficient to prevent a long-distant effect on the positron's spin wavefunction 
due to measurements on the spin of the electron. 
Aspect (ii) of the EPR paradox cannot happen without violating quantum-mechanical unitarity. 

As a side comment, it is possible for the linear combination $a \ket{\uparrow} + b \ket{\downarrow}$
to become either only up spin or down spin during a measurement process without violating unitarity 
by having ${\ket{\downarrow}}$ evolve to ${\ket{\uparrow}}$ through {\it physical} interactions: 
${a \ket{\uparrow} + b \ket{\downarrow}} \to {a \ket{\uparrow} + b' \ket{\uparrow}} 
= {(a + b') \ket{\uparrow}}$ where  ${\langle b' \vert b' \rangle } = {\langle b \vert b \rangle }$ 
and  ${\langle a \vert b' \rangle} = 0$. 
The Converse Collapse Statement still holds even if there is a unitarity-preserving spin-${1 \over 2}$ measurement 
that mimics spin wavefunction collapse through physical effects 
(in the sense that if the spin is measured to be up then 
at some point during the measuring process ${a \ket{\uparrow} + b \ket{\downarrow}} \to  {(a + b') \ket{\uparrow}}$, 
that is, the down component of spin is not destroyed but gets flipped).
However, a unitarity-preserving spin-${1 \over 2}$ measurement 
still cannot render the spin part of the wavefunction as $\ket{\uparrow}_-\ket{\downarrow}_+$ or $\ket{\downarrow}_-\ket{\uparrow}_+$ 
without possibly violating angular momentum if the initial state is an eigenstate of angular moment.

Although not rigorous, one can argue that spin wavefunction collapse cannot happen on physical grounds 
within first-quantized quantum mechanics. 
For example, if Stern-Gerlach\cite{SternGerlach} is used to measure the spin of a spin-${1 \over 2}$ object, 
then the form of the wavefunction after passing through the gradient magnetic field is 
$a \Psi_{\uparrow} (\vec x) \ket{\uparrow}  + b \Psi_{\downarrow} (\vec x)  \ket{\downarrow}$, 
where the support of $\Psi_{\uparrow} (\vec x)$ (respectively, $\Psi_{\downarrow} (\vec x)$) 
is largely for $z > 0$ (respectively,  $z < 0$) 
when the magnetic field is oriented in the $z$ direction.
A detection screen is then used to locate the object. 
If the object is detected to be above (respectively, below) the $z = 0$ plane, 
then the spin is considered to be up (respectively, down). 
Consider the case when the spin is measured to be up.
There is nothing within typical first-quantized processes capable of destroying $\Psi_{\downarrow} (\vec x)$.
If, for example, the object forms a bound state with entities in the $z > 0$ region of the detection screen 
then the component $ \Psi_{\downarrow} (\vec x)$ could not participate in the formation of this state 
without ruining the spin measurement since there would be no reason 
for the bound state to form above the screen with probability $|a|^2/(|a|^2+|b|^2)$. 
Typical nanometer forces also do not ``draw in'' such a`` faraway'' piece of the wavefunction. 
Indeed, linearity, which is a fundamental property of first-quantized quantum mechanics, implies that the final state 
will be of the form $a \Psi'_{\uparrow} (\vec x, \vec s)  + b \Psi'_{\downarrow} (\vec x, \vec s)$ 
where $\Psi'_{\uparrow}$ is a bound state in the $z > 0$ region and $\Psi'_{\downarrow} (\vec x, \vec s)$ 
is a bound state in the $z < 0$ region. 
In other words, the final states is a linear superposition of two different bound states. 
We indicate that spin here as $\vec s$ because when the object interacts with the degrees of freedom 
of the $z>0$  (respectively, $z<0)$ detection screen, the spin is unlikely to remain as $\ket{\uparrow}$ (respectively, $\ket{\downarrow}$.
Even before the object strikes the screen, 
a typical gradient magnetic  field of a Stern-Gerlach device causes the spin to be twisted,\cite{SStwistedSpin}
meaning that at a fixed time, the spin points in a helix-like set of directions in spin space. 
Thus, it is very unlikely that the spin ever becomes purely up or purely down 
as is required by wavefunction collapse. 
This is the situation for the spin measuring device described in the first two paragraphs of Appendix A: 
If the spin of the object to be measured enters as $a \ket{\uparrow}  + b \ket{\downarrow}$, 
then when the spin-flipping takes place, the object's spin always ends up as $\ket{\uparrow}$. 
Hence, if the spin is measured to be down, then it never becomes $\ket{\downarrow}$ 
at any point during the measuring process. 
Since the spin measuring device in Appendix A does not violate any principles of quantum mechanics, 
the final state is a linear superposition of $a \Psi'_{\uparrow}  + b \Psi'_{\downarrow}$ due to linearity, 
where $\Psi'_{\uparrow}$ (respectively, $ \Psi'_{\downarrow}$) is a wavefunction 
involving an ``up'' (respectively, ``down'') signal. 
Section \ref{SecondQuantization} provides a resolution of the Measurement Problem, 
that is, why there appears to be a unique signal generated by measuring devices.

We have shown that an EPR paradox cannot arise in ordinary first-quantized quantum mechanics 
for the entangled  spin-${1 \over 2}$ system 
because the paradox depends on spin wavefunction collapse, 
and in unitary quantum mechanics, such a collapse cannot occur. 

The first experiments\cite{FreedmanClauser,Aspect1981,Grangier1982,Dalibard1982,OuMandel,ShihAlley} involving the EPR paradox
were conducted using entangled photons in the context of testing Bell's inequality\cite{Bell} 
or the Clauser-Horne-Shimony-Holt\cite{CHSH} version of it.
Compared to the above discussion involving spin-$1 \over 2$ particles, 
photon polarizations play the role of spin up and spin down.
Two photons with polarization structure of the form ${( \ket{H}_- \ket{V}_+ -  \ket{V}_- \ket{H}_+ )/\sqrt{2}}$, 
where $H$ and $V$ indicate horizontal and vertical polarization states, 
were separated by a significant distance.  
These experiments and others since the 1980's consistently found the two polarization signals 
were correlated: when the measurement of one device yields $H$, the other one yields $V$. 
Furthermore, if the axis of measurement for polarization for the photon on the right is 
at an angle $\theta$ with respect to the axis of measurement for the polarization for the photon on the left, 
then if the left device yields $H$ (respectively, $V$), 
then the right device measures $V$ (respectively, $H$) $cos^2(\theta)$ of the time and 
$H$ (respectively, $V$) $sin^2(\theta)$ of the time within experimental uncertainties, 
just as one would expect from quantum mechanics from the term $ \ket{H}_- \ket{V}_+$ (respectively, $\ket{V}_- \ket{H}_+$).
Since these experiments use photons, 
second-quantization processes are involved 
and therefore they do not test the EPR paradox in first-quantized quantum mechanics.

There have been a few experiments testing the spin-$1 \over 2$ system.\cite{Hensen,Rosenfeld}
However, these experiments still rely on photons. 
For example, in a 2015 experiment,\cite{Hensen} the entanglement of two spin-${1 \over 2}$ electrons 
was achieved using a clever version of entanglement swapping.\cite{BarrettKok}
The swapping is actually carried out using photons. 
After the electrons had been separated by 1.3 kilometers, 
the spin of each electron was measured using spin-dependent fluorescence, 
again using photons. 
Hence, second-quantized processes were involved throughout the experiment. 
The spins of the two electrons were found to be opposite within experimental errors. 
If the axis of quantization for spin for the electron on the right is 
at an angle $\theta$ with respect to the axis of quantization for spin for the electron on the left, 
then if the left device yields $\ket{\uparrow}$ (respectively, $\ket{\downarrow}$), 
then the right device measures $\ket{\downarrow}$ (respectively, $\ket{\uparrow}$) $cos^2(\theta/2)$ of the time and 
$\ket{\uparrow}$ (respectively, $\ket{\downarrow}$) $sin^2(\theta/2)$ of the time within experimental uncertainties, 
just as one would expect from quantum mechanics from the term 
$\ket{\uparrow}_-\ket{\downarrow}_+$ (respectively, $\ket{\downarrow}_-\ket{\uparrow}_+$), 
a result that follows from 
$\ket{\uparrow} = \cos({\theta \over 2}) \ket{\uparrow}_\theta + \exp(i \beta)  \sin({\theta \over 2}) \ket{\downarrow}_\theta$ 
and 
$\ket{\downarrow} =  - \exp(-i \beta)  \sin({\theta \over 2}) \ket{\uparrow}_\theta +  \cos({\theta \over 2}) \ket{\downarrow}_\theta$ 
for some $\beta$ and where $ \ket{ }_\theta$ indicates a spin-${1 \over 2}$ basis 
at an angle $\theta$ with respect to the one for the electron on the left.

In summary, the EPR paradox has not yet been tested in purely first-quantized quantum mechanics
for the entangled spin-${1 \over 2}$ system. 
In Section \ref{SecondQuantization}, we discuss the some of the issues 
concerning wavefunction collapse and its timing in quantum field theory. 

\section{Measuring Spin-${1 \over 2}$}
\label{SGmeasurement}

If wavefunction collapse does not happen, then the Converse Collapse Statement holds. 
In this section, we show that, in general, many measurements are needed to establish the value of the spin of a spin-${1 \over 2}$ object. 
We shall need to assume that the concept of measurement is possible in first-quantized quantum mechanics 
and that a device exists to perform the measurement.

The measuring device needs to be equipped with a directional ability, 
that is, when it is oriented in the $\vec n$-direction, it is capable of measuring up and down spin in this direction.
Let us begin with $\vec n$ pointing in the $z$-direction.
If the spin of the object is arranged to either point up or point down 
(and this can be decided randomly and unbeknownst to the experimentalist) 
but is not a linear combination of these two states, then a {\it single} measurement suffices to determine the spin. 
If the spin part of the wavefunction of the object is $\ket{\uparrow}$, 
the device will produce  an ``up'' signal and the experimentalist knows the spin is up. 
However, no wavefunction collapse is needed in this situation: 
The spin before entering the device is $\ket{\uparrow}$, 
and so when it is observed to be up, the spin part of the wavefunction did not need to collapse. 
The same is true if the spin is down: 
the spin wavefunction enters the device as $\ket{\downarrow}$, 
it is observed to be down and therefore remains unchanged 
at least until the object begins interacting with the device. 

When the spin part of the wavefunction of the object is a linear combination of up and down spin, 
that is, it is of the form $a\ket{\uparrow} + b\ket{\downarrow}$ with $a \ne 0$ and $b \ne 0$, 
then a single measurement is insufficient to determine the spin of the object. 
The device is capable of producing an ``up'' signal or a ``down'' signal 
but not an output of the numbers ``a'' and ``b''. 

When the spin is of the form $a\ket{\uparrow} + b\ket{\downarrow}$, 
there is a direction $\vec n$ of quantization in spin space in which the spin points up.\footnote{When the spin is 
of the form $(\alpha + i\gamma)\ket{\uparrow} + (\beta + i \delta)\ket{\downarrow}$, 
where $\alpha$, $\beta$, $\gamma$ and $\delta$ are real numbers, 
this direction $\vec n$ in $3$-space is given by 
$\vec n = (2(\alpha \beta + \gamma \delta), 2(\alpha \delta - \beta \gamma), 
  \alpha^2 - \beta^2 + \gamma^2 - \delta^2)/(\alpha^2 + \beta^2 + \gamma^2 + \delta^2)$.
}
To measure the spin with the device, one must orient it in this direction. 
This requires generating the incoming object with the same spin state each time and performing repeated measurements. 
Although it should be obvious how to do this, we present the procedure for completeness 
and because it emphasizes the need for multiple measurements.

After performing many measurements with the device oriented in the $z$-direction, 
one notes the number of ``up'' signals $N_{up}$  and the number of ``down'' signals $N_{down}$. 
Then the magnitudes of $a$ and $b$ (up to experimentally uncertainty) can be obtained from
\begin{equation}
   \vert a \vert^2 = \frac{N_{up}}{N_{total}} \  ,  \ \  \vert b \vert^2 = \frac{N_{down}}{N_{total}} 
   \  ,  \ \ where \ \ N_{total} = N_{up} + N_{down}  
\, .
\label{SpinComponentMagnitude}
\end{equation}
It then follows that the spin of the object is located at an angle $\theta$ with respect to the $z$-axis, where
\begin{equation}
   \tan{({\theta \over 2})} =  { {\vert b \vert} \over  {\vert a \vert}}  = \sqrt{ {N_{down}} \over {N_{up}}}
\, .
\label{SpinAngle}
\end{equation}
If $\theta = 0$ (respectively, $\pi$), then the spin points in the positive (respectively, negative) $z$-direction. 
If $\theta$ is neither $0$ nor  $\pi$, then there is a circle of possibilities: 
the ring in polar coordinates located at the angle $\theta$ of Eq.(\ref{SpinAngle}) with respect to the $z$-axis. 
One can then orient the device at an angle $\theta$ with respect to the $z$-direction (or even at another angle) 
and perform repeatedly measurements 
until the number of ``up'' signals $N'_{up}$  and the number ``down'' signals $N'_{down}$ is accurately determined. 
Here, the primes indicate spin measurements in the direction of the new axis. 
By using Eq.(\ref{SpinAngle}), one can determine that the spin points at an angle $\theta'$ with respect to the new axis. 
This produces a second circle of possibilities for the orientation of the spin. 
The intersection of these two circles generates two points at most. 
In general, the device needs to be oriented in one of these two directions 
and a few more measurements are required to determine the direction of the spin of the object. 

When this direction is $\vec n = (n_x, n_y, n_z)$ where $n_x^2 + n_y^2 + n_z^2 = 1$, 
the spin of the object (up to an overall phase) is equal to
\begin{equation}
   a\vert \uparrow \rangle + b\vert \downarrow \rangle = \sqrt{\frac{(1+n_z)}{2}} \vert \uparrow \rangle + 
   (n_x + i n_y ) \frac{\sqrt{1-n_z}}{\sqrt{2} \sqrt{n_x^2 + n_y^2} } \vert \downarrow \rangle
\, .
\label{SpinSolution}
\end{equation}
In Eq.(\ref{SpinSolution}), we have adjusted the overall phase so that the coefficient $a$ is real.

That the above procedure is a correct method for measuring the spin of a spin-$1 \over 2$ object 
using the device should be obvious. 
However, the message that emerges from the discussion of this procedure is that 
a single measurement cannot determine the spin of the object. 
It takes many measurements to establish Eq.(\ref{SpinSolution}); 
each measurement provides limited information and 
only when combined with the results of many other measurements can the spin be determined. 

To summarize, 
when an ``up'' (respectively, ``down'') signal occurs in a single measurement, 
the spin of the object does not necessarily point up (respectively, down), 
and the experimentalist cannot be sure in which direction the spin points at any moment during the measuring process. 
The uncertainty at the microscopic level of quantum mechanics is transmitted to the macroscopic level 
in that the experimental signal (be it ``up''  or  ``down'') is uncertain 
in conveying the status of the spin to the experimentalist. 

The above picture is in contrast to the wavefunction collapse one.
In that picture, an ``up'' signal (respectively, ``down'' signal)
is considered to indicate that the spin is definitely up (respectively, down). 
Given the connection between probability and the wavefunction, 
the spin wavefunction must collapse to $\ket{\uparrow}$ (respectively, $\ket{\downarrow}$) to reflect this certainty.

In first-quantized quantum mechanics, 
it may be that the device is incapable of producing a signal when 
it is not oriented in the direction of the spin, that is, 
when the spin of the object is of the form 
$a\ket{\uparrow} + b\ket{\downarrow}$ with $a \ne 0$ and $b \ne 0$.
In this case, it is still possible to measure the spin of the object but the procedure is laborious. 
One would need to randomly orient the device in various directions 
until a signal is produced. 
One would then know that the spin points in this direction 
and the ``up''  or ``down'' signal would indicate whether the spin is up or down. 

It is interesting to note that
repeated measurements are also needed if collapse of the spin part of the wavefunction takes place: 
Although one knows for certain what is the spin at the time of the measurement, 
information about the spin before the measurement is lost if the spin wavefunction is of the form 
$a\ket{\uparrow} + b\ket{\downarrow}$ with $a \ne 0$ and $b \ne 0$.
So the procedure for determining the spin in the case when collapse of the spin part of the wavefunction is possible 
is the same as above: 
multiple measurements and difference orientations of the device are needed. 
Without wavefunction collapse, the measurement itself is uncertain;
with wavefunction collapse, the measurement is certain but information about the original spin state 
is lost in the measuring process.

\section{Measurements of Quantum Systems in General}
\label{generalmeasurement}

Consider a macroscopic measuring device that is capable of 
detecting up to $M$ different quantum states, $s_1$, $s_2$, $\dotsc$, $s_M$ normalized to $1$: $ \langle s_i , s_i \rangle = 1$. 
In principle, these states need only to be distinct, which can be represented mathematically as $ \langle s_i , s_j \rangle < 1$ for $i \ne j$; 
however, some of the statements below are valid only if the states are orthogonal. 
Hence, for convenience, we assume 
\begin{equation}
 \langle s_i , s_j  \rangle = \delta_{ij}
\, .
\label{DistinctStates}
\end{equation}

We require the measuring device to be ``good in the first sense'', the definition of which is as follows:
If the incoming state (that is, the state to be measured) is purely $s_i$ 
then the device produces the signal $S_i$.
We represent this by 
\begin{equation}
   s_i \to S_i
\, .
\label{ProperSignalling}
\end{equation}
In the case of  spin-${1 \over 2}$ discussed in the previous section, 
there are two states ($M=2$) and two signals $S_i$,  $i = 1$, $2$.
Note that no collapse of the wavefunction is needed for the situation corresponding to  Eq.(\ref{ProperSignalling}): 
The signal $S_i$ indicates that the state has to be $s_i$ 
and indeed it is $s_i$ before being measured by the device.  

Let us first assume that collapse cannot happen for any wavefunction.  
Let the incoming state $s_0$ be a  linear combination of the $s_i$, that is, 
\begin{equation}
  s_0 = \sum_i { a_i s_i }
\, ,
\label{LinearCombinationInitialState}
\end{equation}
where $a_i$ are complex numbers whose absolute squares sum to $1$.
Linearity, which is a fundamental property of quantum-mechanics,
says that if Eq.(\ref{LinearCombinationInitialState}) holds at time $t_0$ 
then one may evolve each component wavefunction $s_i$ from time $t_0$ to a later time $t_f$ 
independently of the other components and then perform the sum in Eq.(\ref{LinearCombinationInitialState}) to obtain $s_0$ at a later time. 
This follows because the evolution operator in Eq.(\ref{UnitaryOperator}) is linear: 
\begin{equation}
\begin{aligned}
                    &  s_0 (t_0) =  \sum_i { a_i s_i   (t_0) } \implies \\ 
  s_0 (t_f) =  & \  U(t_f, t_0) s_0(t_0) =  U(t_f, t_0) \sum_i { a_i s_i (t_0) } =   \\
                    & \sum_i { a_i   U(t_f, t_0) s_i (t_0) } =   \sum_i { a_i  s_i (t_f) }
\, .
\label{Linearity}
\end{aligned}
\end{equation}
In terms of the Hamiltonian $H$, linearity is expressed as $ H (\sum_i a_i s_i) = \sum_i a_i (H s_i)$.
Linearity then implies that the wavefunction of the measuring device 
must eventually be a linear superposition of states, each of which contains one of the signals $S_i$: 
\begin{equation}
  \sum_i a_i s_i ( \{x\}, \{e\}, t_0) \to \sum_j   a_i s_i ( \{x\},  \{e\}, t_f | S_i ) \ \ \ (Linear \ Picture )
\, .
\label{LinearPicture}
\end{equation}
Here, the arrows indicate the evolution of the wavefunction, 
$t_0$ (respectively, $t_f$) is a time before (respectively, after) which the measurement has taken place, 
and $ s_i ( \{x\}, \{e\}, t_f | S_i )$ is a wavefunction in which the signal $S_i$ is present. 
In Eq.(\ref{LinearPicture}), 
we explicitly show the variable dependence of the states $s_i$: 
they depend on variables $\{x\}$ associated with the object to be measured and 
all other variables $\{e\}$ not associated with this object including those of the environment
and the measuring apparatus. 
Note that $ \{x\}$ and $\{e\}$ represent sets of variables. 
Initially, it may be possible to factorize the wavefunction as $s_i ( \{x\}, \{e\}) = \tilde s_i ( \{x\}) \Psi_i (\{e\})$ 
but when the object to be measured interacts with the environment, the apparatus or both, 
this factorization is almost surely not possible. 

In the picture in which measurement is linked to wavefunction collapse, 
it is {\it assumed} that if the device produces the signal $S_i$ 
then the incoming state must be $s_i$ when measured. 
This then requires $s_0$ in Eq.(\ref{LinearCombinationInitialState}) to become $s_i$ at some point in time: 
\begin{equation}
  \sum_j a_j s_j \to s'_i \to S_i
\, ,
\label{WavefunctionCollapse}
\end{equation}
where $ s'_i$ is the evolution of $s_i$ at a later time. 
In the wavefunction collapse picture, this is achieved 
by having $\vert a_i \vert \to 1$ and $\vert a_j \vert \to 0$ for $j \ne i$. 
Uncertainty of quantum mechanics at the microscopic scale is lost at the macroscopic scale.

In the wavefunction collapse picture, Eq.(\ref{LinearPicture}) is replaced by 
\begin{equation}
  \sum_i a_i s_i (\{x\}, \{e\}, t_0) \to s_j (\{x\}, \{e\}, t_m) \to  s_j (\{x\}, \{e\}, t_f | S_j ) \ \ \ (Collapse \ Picture )
\, .
\label{CollapsePicture}
\end{equation}
Eq.(\ref{CollapsePicture}) produces what is observed experimentally: 
Each time the experiment is conducted, only one signal $S_j$ occurs 
as represented by the last term $s_j (\{x\}, \{e\}, t_f | S_j )$ in  Eq.(\ref{CollapsePicture}).
The collapse of the wavefunction occurs at a certain point in time $t_m$
when the components $s_i$ for $i \ne j$ are destroyed and the entire wavefunction becomes $s_j$ up to an overall phase, 
as is indicated in the middle term in Eq.(\ref{CollapsePicture}). 
The wavefunction collapse picture violates unitarity, linearity and it is necessary to assume that 
the measuring devices are ``good in the second sense''
in that the probability of having the state with the signal $S_j$ is $|a_j|^2$. 

The linear picture necessarily has generalized Schr\"odinger Cats, 
that is, states that are a linear superposition of macroscopically seemingly incompatible situations. 
The ``traditional'' Schr\"odinger Cat\cite{SchrodingerCat} can be obtained by having one of the signals 
in Eq.(\ref{LinearPicture}) connected to the release of poison. 
Wavefunction collapse avoids Schr\"odinger Cats 
because it assures that the animal is {\it either} alive {\it or} dead. 

With the same source (Eq.(\ref{LinearCombinationInitialState})) and repeated measurements, 
sometimes the experimentalist observes $S_1$, sometimes $S_2$, etc. 
After the experiment is repeated $N_{total}$ times, 
then one deduces the magnitudes of the $a_j$ coefficients up to experimental uncertainties from
\begin{equation}
  \vert  a_j \vert ^2 = {{N_j} \over {N_{total}}}
\, ,
\label{ajDetermination}
\end{equation}
where $N_j$ is the number of times the signal $S_j$ was observed during $N_{total}$ measurements. 

The situation at this point is analogous to the case of measuring spin-$1 \over 2$ with a device 
that is oriented only in the $z$-direction. 
To determine the real and imaginary parts of the $a_j$, 
new devices, which are capable of measuring different linear combinations of the $s_i$ must be used. 
The analogue for the spin-$1 \over 2$ case of Section \ref{SGmeasurement} 
is pointing the device in a different direction. 

In a single measurement, the experimentalist cannot be sure 
that the incoming state is $s_i$ when the signal on the device is observed to be $S_i$. 
In the linear picture, the uncertainty of quantum mechanics at the microscopic level 
is transferred to uncertainty at the macroscopic level.
From the analysis in this section emerges the following {\bf Quantum Measurement Rule}: 
{\it If wavefunction collapse is needed ``to explain'' an experimental result, 
then multiple measurements are necessary to determine the quantum state,
but if a measurement does not need wavefunction collapse to explain it then a single measuring event suffices.} 
Alternatively, in terms of the measuring capability of the device described in the first paragraph of the section, 
which involves the states $s_i$, 
multiple measurements are needed whenever the incoming state is a linear combination of the $s_i$.

\section{Quantum Field Theory}
\label{SecondQuantization}

The discussion and analysis up to now has been within the context of ordinary first-quantized quantum mechanics
with the key result being that wavefunction collapse cannot happen in this theory. 
In this section, we discuss related topics in quantum field theory. 
We warn the reader that this section involves some speculations. 

The primary difference between first-quantized quantum mechanics and second-quantized quantum mechanics 
is that the latter includes the possibility of having particles produced or destroyed 
and that Fock spaces are needed to account for this.
A secondary difference is that in quantum field theory there are Feynman diagrams involving loops
that have no counterpart in ordinary first-quantized quantum mechanics; 
however these can often be incorporated into corrections and effective potentials in a first-quantized approach. 
Hence, the use of ordinary quantum mechanics 
is expected to be reasonably accurate between particle creation and destruction events.
Given this, 
wavefunction collapse, {\it if it is possible}, 
most likely occurs when particles are created or destroyed. 

Wavefunction collapse involves a drastic, rather sudden change in a wavefunction.
This is also true for particle production and destruction processes. 
Our understanding of these processes is limited and entirely based on perturbation theory. 
Hence, one cannot completely exclude the possibility
that linearity and unitarity in quantum field theory are violated non-perturbatively and that wavefunction collapse happens
during the production and destruction of particles.

Consider the situation in which a charged particle passes near an object which has the capability of stimulating the former to emit a photon 
and that later, when the particle is far away, no further stimulated emission of a photon is possible. 
For simplicity, we exclude the case of multiple photon emission. 
Let $\Psi(\{ x \} )$ be the wavefunction before the particle has reached the region of the object. 
Here, ${x}$ stand for all the degrees of freedom of the system. 
Then at a sufficiently later time, the form of the wavefunction is $\Psi_\emptyset(\{ x \} ) + \Psi_\gamma(\{ x, x_\gamma \} )$
where $\Psi_\gamma$ (respectively, $\Psi_\emptyset$) represents the situation in which the photon is (respectively, is not) emitted. 
Note that $\Psi_\gamma$ involves additional degrees of freedom for the photon, 
which we represent by $\{  x_\gamma \}$, and that
$\bracket {\Psi_\gamma}{\Psi_\gamma}$ (respectively, $\bracket {\Psi_\emptyset}{\Psi_\emptyset}$) 
is the probability that the photon is (respectively, is not) emitted. 

When a system initially described by a wavefunction $\Psi(\{ x\}, t_0 )$ 
involves particle production or destruction, 
the final wavefunction subsequently becomes 
\begin{equation}
     \Psi(\{ x\}, t ) =  \Psi_A(\{ x_A \} , t ) + \Psi_B(\{ x_B \} , t ) 
\, ,
\label{DistinctiveQuantumStates}
\end{equation}
where $ \Psi_A(\{ x_A \})$ and $\Psi_B(\{ x_B \})$ are in different Fock spaces. 
In the example of the previous paragraph, one can take $\Psi_A(\{ x_A \}) = \Psi_\emptyset(\{ x \} )$
and $\Psi_B(\{ x_B \} ) = \Psi_\gamma(\{ x, x_\gamma \} )$.
If linearity holds in quantum field theory,
then even if $A$ and $B$ initially represent ``microscopic differences'', 
$A$ and $B$ will eventually lead to a generalized Schr\"odinger Cat 
if the degrees of freedom associated with the microscopic differences 
interact with degrees of freedom of a macroscopic object, 
and this is reasonably likely to happen given enough time.  
We call such a ``Cat'' a {\it Fock-Space Schr\"odinger Cat}.
It is also not too difficult to arrange $A$ and $B$ to initially correspond to a microscopic difference 
that leads to the traditional Schr\"odinger Cat in which $ \Psi_A$ involves a wavefunction associated with a dead cat 
and $ \Psi_B$ involves a wavefunction associated with a live cat. 
It is often said that this is a paradox in quantum mechanics 
because it corresponds to a situation in which the cat can be alive {\it and} dead at the same time. 
Actually, a wavefunction of the form of Eq.(\ref{DistinctiveQuantumStates}) corresponds 
to a situation in which $A$ happens with probability $\bracket {\Psi_A}{\Psi_A}$ {\it or} 
$B$ happens with probability $\bracket {\Psi_B}{\Psi_B}$ 
because $\Psi_A$ and $\Psi_B$ are orthogonal. 
A generalized Schr\"odinger Cat just corresponds to a wavefunction $\Psi (\{x\})$
with a ``very bifurcated'' probability distribution $|\Psi (\{x\})|^2$ for a macroscopic object. 
A living organism always has some distribution in its constituent variables at the nanometer scale. 
What is unusual about a generalized Schr\"odinger Cat is that the distribution 
has separated into two very distinct pieces. 
There is nothing within quantum mechanics that prohibits this from happening 
and indeed in unitary quantum mechanics (which does not allow wavefunction collapse to occur), 
this has to sometimes happen.
However, a person who finds this situation disturbing will be tempted to invoke wavefunction collapse, 
thereby eliminating one of the two components in the linear superposition of Eq.(\ref{DistinctiveQuantumStates}).

We now {\it assume} that wavefunction collapse occurs during particle production or destruction 
as A. V. Melkikh has suggested\cite{Melkikh} and investigate the consequences. 
When these events happen, the wavefunction, 
which would have been $\Psi_A(\{ x \} ) + \Psi_B(\{ x \} )$ without wavefunction collapse, 
changes in the following manner: 
\begin{equation}
\begin{aligned}
    Wa&vefunction \ Collapse:    \\
  \Psi_A(\{ x \} )  +  \Psi_B&(\{ x \} )  \to {1 \over {\bracket{\Psi_A}{\Psi_A}}} \Psi_A(\{ x \} ) \ \ with \ probability \  \bracket{\Psi_A}{\Psi_A}  \\
\label{FockSpaceWavefunctionCollapse}
  or  &  \\
  \Psi_A(\{ x \} )  + \Psi_B&(\{ x \} ) \to  {1 \over {\bracket{\Psi_B}{\Psi_B}}}  \Psi_B(\{ x \} ) \ \ with \ probability \  \bracket {\Psi_B}{\Psi_B}  
\, .
\end{aligned}
\end{equation}

Let us now consider the EPR paradox when wavefunction collapse is possible as in Eq.(\ref{FockSpaceWavefunctionCollapse}) 
during particle creation and destruction events. 
Since most of the tests of the EPR paradox involve Bell inequality experiments\cite{BellTestExperiments} with photons, 
we shall focus on this case. 
These experiments are very similar to the entangled spin-${1 \over 2}$ system 
discussed in the first part of this article 
with photon polarizations playing the role of spin up and spin down. 
Indeed, all one needs to do is to replace $\ket{\uparrow}$ by $\ket{H}$ and $\ket{\downarrow}$ by $\ket{V}$. 
The entangled polarizations in the singlet state have the structure
\begin{equation}
\ket{singlet} = {1 \over {\sqrt{2}}} ( \ket{H}_{-} \ket{V}_{+} -  \ket{V}_{-} \ket{H}_{+})
\, ,
\label{PolarizationSingleState}
\end{equation}
where the minus (respectively, plus) subscript indicates the photon moving to the left (respectively, right). 
The setup is still given in Figure \ref{fig:EPRsetup} with $e^-$ and $e^+$ replaced by photons 
and with polarization measuring devices replacing the spin-${1 \over 2}$ measurers. 
The polarization measuring device first spatially separates the horizontal and vertical components of the photon wavefunction
(in the analogous way in which a $z$-directed gradient magnetic field in Stern-Gerlach\cite{SternGerlach} 
separates up spin from down spin for a spin-${1 \over 2}$ object).
The photon is then detected and if this detection occurs in the region where the horizontal (respectively, vertical) polarization component 
is propagating then one says that the photon has been measured to be in the horizontal (respectively, vertical) polarization state. 
This is analogous to the detection screen in a Stern-Gerlach method for measuring spin-${1 \over 2}$ for which
a flash above the $z=0$ plane indicates that the spin is in an up state 
and a flash below the $z=0$ plane indicates that it is in a down state.

Let us now re-examine the arguments ruling out wavefunction collapse for first-quantized quantum mechanics 
in Sections \ref{AngularMomentumViolation} - \ref{EPRresolution} 
to see if they apply to the EPR tests for the Bells inequality experiments that use entangled polarizations of photons. 
The argument concerning the violation of the Heisenberg uncertainty principle 
fails because the little group determining photon polarizations is $O(2)$ and hence abelian. 
It is not clear how the gedanken experiment 
leading to violation of angular momentum might extend to the entangled photon case. 
However, even if it does, there is a way to avoid the problem: 
If the production or destruction of photons leads to a violation of a conservation law 
then the process cannot happen; the signal would be prohibited from happening 
and wavefunction collapse would not occur. 
Of course, the angular momentum violation argument against wavefunction collapse in Section \ref{AngularMomentumViolation} 
does not apply to the Bell inequality experiments that have been performed 
because the initial state is not an eigenstate of angular momentum.

The strongest argument against wavefunction collapse in first-quantized quantum mechanics is unitarity. 
Our understanding of quantum field theory, and in particular quantum electrodynamics, 
is almost entirely derived from perturbation theory. 
It is known, however, that one cannot define the theory by summing all  terms in perturbation theory:
the series is asymptotic.\cite{DysonAsymptotic}
Since unitarity is verified at each order in perturbation theory, 
a non-perturbative violation of unitarity would need to occur. 
One can only speculate on how this might happen: 
In first-quantized quantum mechanics, 
mathematically rigorous definitions of the Feynman path integral 
(which is a way of formulating the evolutionary unitary operator in Eq.(\ref{UnitaryOperator})) 
can be established.\cite{SSFeynmanIntegrals}
This is not the case for quantum field theory and for quantum electromagnetism in particular,
where the functional integral -- the quantum field theory analog of the Feynman path integral -- 
is the most common way to represent the evolutionary operator. 
The chiral anomaly and other anomalies provide examples in which a property, such as chiral symmetry, 
of a quantum field theory naively should hold 
but a mathematical obstruction arises that leads to its violation.\cite{AdlerChiralAnomaly,BellJackiw}
So, perhaps a similar issue arises for unitarity. 
Quantum field theory requires the regularization of infinities, which are eventually eliminated by the renormalization program. 
Regardless of the process (electron-electron scattering, electron-positron annihilation into photons, etc.) 
the same renormalization procedure works in perturbation theory. 
However, given two initial wavefunctions $\Psi_A$ and $\Psi_B$, it is possible to imagine that  
non-perturbatively the renormalization procedures for $\Psi_A$ and the one for $\Psi_B$ might be different.
Then the evolution of $\Psi_A$ and the evolution of  $\Psi_B$ would not be sufficient 
to determine the evolution of  $a \Psi_A + b \Psi_B$, where $a$ and $b$ are constants. 
Violation of linearity implies violation of unitarity. 
All the above is speculation but indicates that we cannot rigorously rule out the unitarity argument 
against wavefunction collapse in quantum field theory. 

Now consider the case of a moving measuring device as in Figure \ref{fig:setupMoving} 
for the entangled photon case. 
If wavefunction collapse occurs instantly, 
then, in the rest frame of the left experimentalist, 
the measuring device on the left causes photon polarization collapse 
before the measurement on the right has occurred. 
The same is also true in the rest frame of the right experimentalist. 
In her rest frame, she measures the photon polarization before the experimentalist on the left has performed his measurement. 
See Figures \ref{fig:differentReferenceFrames}(b) and \ref{fig:differentReferenceFrames}(c). 
It would then appear that not only does ``common sense'' special relativity 
need to be violated but quantum field theory has to ``know the future in advance''
to explain why the collapse of the wavefunction to the component involving $\ket{H}_- \ket{V}_+$ (respectively, $ \ket{V}_- \ket{H}_+$) 
on the left can be coordinated with the collapse of the wavefunction 
on the right to the component involving $\ket{H}_- \ket{V}_+$ (respectively, $ \ket{V}_- \ket{H}_+$). 
Of course, the Bells inequality experiments have not been performed for the relativistically moving detector case
of Figure \ref{fig:setupMoving}.
It is possible, but unlikely in my opinion, that moving measuring devices would make a difference in the outcome of these experiments.
Accepting this and assuming that wavefunction collapse happens during photon production and destruction, 
one concludes that the collapse cannot happen instantaneously 
given the fact that quantum field theory respects special relativity, is causal and obeys locality.  

It is interesting to explore what happens in a model of wavefunction collapse 
in which  ``common sense'' causality is required.
Consider the entangled photon polarization experiments 
for which  the situation in Figure \ref{fig:differentReferenceFrames}(a)
is relevant because both measuring devices and source have a common rest frame. 
Suppose that only one polarization measurement takes place at the space time point $M_L = ({\bf y}_{L0} , t_{L0})$ 
at which particle destruction or production takes place. 
If wavefunction collapse does not immediately happen and linearity holds, 
then the wavefunction initially changes from 
\begin{equation}
\begin{aligned}
 {1 \over {\sqrt{2}}} (\Psi_{HV} (\{ x_L \}, \{ x_R \}  , t) \ket{H}_{-} \ket{V}_{+} &- \Psi_{VH} (\{ x_L \}, \{ x_R \} ,  t) \ket{V}_{-} \ket{H}_{+})   \ \ to \\
 {1 \over {\sqrt{2}}} (\Psi'_{HV} (\{ x'_L \}, \{ x_R \} , t) \ket{V}_{+} &- \Psi'_{VH} (\{ x'_L \}, \{ x_R \} ,  t) \ket{H}_{+}) 
\ 
\label{OnePolarizationMeasurement}
\end{aligned}
\end{equation}
when the measurement is made.
Here, the $\{ x \}$ contain not only the spatial coordinates of all relevant degrees of freedom 
(those of the two photons, the constituents of the experimental equipment, the experimentalists themselves, etc.) 
but also other degrees of freedom such as spins and polarizations,  
and the subscript $L$ indicates those degrees of freedom associated with the region 
to the left in Figures  \ref{fig:differentReferenceFrames}(a) and \ref{fig:LightCone} (with $e^+$ and $e^-$ replaced by photons). 
During the measuring process, 
the spatial dependence of the two polarization terms evolve differently 
(and hence the two functions $\Psi_{HV}$ and $\Psi_{VH}$) and,  
at the point at which the measurement is made, 
the degrees of freedom may change (hence the prime on ${ x'_L }$). 
Since, after the polarization has been measured, the photon moving to the left might have been destroyed or its polarization changed, 
one cannot guarantee that $\Psi_{HV}$ (respectively, $\Psi_{VH}$) will remain proportional to 
$\ket{H}_{-}$  (respectively, $\ket{V}_{-}$)
(and hence its explicit absence in Eq.(\ref{OnePolarizationMeasurement}); 
if the photon still exists, its polarization is included in $\{ x'_L \}$). 
Here, $\Psi'_{HV} (\{ x'_L \}, \{ x_R \} , t)  \ket{V}_{+}$ (respectively,  $ \Psi'_{VH} (\{ x'_L \}, \{ x_R \} ,  t) \ket{H}_{+}$) 
is the wavefunction that would arise in a situation without entanglement in which the initial wavefunction  
is just $\Psi_{HV} (\{ x \}, t) \ket{H}_{-} \ket{V}_{+}$ (respectively, $\Psi_{VH} (\{ x_L \}, \{ x_R \} ,  t) \ket{V}_{-} \ket{H}_{+}$).
When polarization measurements both on the right and on the left are made, 
the wavefunction becomes  
\begin{equation}
 {1 \over {\sqrt{2}}} (\Psi'_{HV} (\{ x'_L \}, \{ x'_R \} , t)  - \Psi'_{VH} (\{ x'_L \}, \{ x'_R \} ,  t) )  
 \ .
\label{TwoPolarizationMeasurements}
\end{equation}
Note that $\Psi'_{HV}$ contains information in it that the photon moving to the left originally involved a horizontal polarization 
while the photon moving to the right involved a vertical polarization. 
This information might be an output reading of an experimental device or an entry in a database. 
An analogous statement holds for $\Psi'_{VH}$.
Let $FLC_{M_L}$ (respectively, $FLC_{M_R}$) be the future light cone that begins at $M_L$ (respectively, $M_R$). 
The intersection of these two light cones is a region of ``common causality''. 
We depict the boundary of this region schematically in Figure \ref{fig:differentReferenceFrames}(a) 
as the ``V'' starting at the space-time point $C$.
Within this region, 
we allow wavefunction collapse to occur: 
\begin{equation}
\begin{aligned}
 {1 \over {\sqrt{2}}} (\Psi'_{HV} (\{ x'_L \}, \{ x'_R \} , t)  - \Psi'_{VH} (\{ x'_L \}, \{ x'_R \} ,  t) )   \to  \\
\Psi'_{HV} (\{ x'_L \}, \{ x'_R \} , t)  \ \ or \ \ - \Psi'_{VH} (\{ x'_L \}, \{ x'_R \} ,  t)  \ .
\label{CoordinatedWavefunctionCollapse}
\end{aligned}
\end{equation}
Actually, the phrase ``within this region'' requires some clarification 
since the wavefunction depends on many variables. 
Roughly speaking, if any variable associated with a polarization measurement signal is within the common causality region 
then we allow wavefunction collapse to occur. 
For example, if the device recording the signals are not moved, 
then wavefunction collapse would happen at the time associated with points $A$ and $B$ 
in Figure \ref{fig:differentReferenceFrames}(a).
If signals were sent electronically to a common location, 
then wavefunction collapse would occur when those signals enter the ``common causality'' region that starts at $C$.

We now show that the above ``delayed'' collapsing model is in agreement with the Bell inequality experiments.\cite{BellTestExperiments}
In our description of the entangled spin-$1 \over 2$ experiment in Section \ref{intro}, 
one step was left out: 
The comparison of the two measurements to verify that the spin at the left and the spin at the right are oppositely oriented. 
In the case of photons, 
the final step is verifying the correlation of $\ket{V}_{-}$ with $\ket{H}_{+}$ and $\ket{H}_{-}$ with $\ket{V}_{+}$.
One way of checking this is to have the experimentalist on the right compare her data with that of the experimentalist on the left. 
Either the experimentalist observes the output directly or the result is recorded in a database and then read by an experimentalist. 
Given that the distance between the measuring devices is no more than a few kilometers
for the experiments that have been conducted up to now, 
the time scale of the delay of wavefunction collapse (the time between $M_L$ and $A$ or between $M_R$ and $B$) 
is the order of $10^{-5}$ seconds or less. 
Hence, by the time\cite{BrainReactionTime} the brains of the experimentalists register the results, 
``coordinated'' wavefunction collapse of Eq.(\ref{CoordinatedWavefunctionCollapse}) will have occurred. 
Therefore, when the two experimentalists compare the results, 
$\ket{V}_{-}$ will be paired with $\ket{H}_{+}$ or $\ket{H}_{-}$ will be paired with $\ket{V}_{+}$.
Another way to compare the polarization outputs
is to (electronically) bring them to a common location.
The polarization results from the two measurements 
are digitally recorded and sent to a common location where they can be placed side by side in a data storage device. 
However, because the transfer of the results to a common location 
must occur at a speed less than that of light, 
the common location will be within the ``common causality'' region that starts at $C$ 
(that is, within both $FLC_{M_L}$ and $FLC_{M_R}$)
and hence within the region where the wavefunction has already collapsed. 
The data entry for one run of the experiment will either contain 
$\ket{V}_{-}$ next to $\ket{H}_{+}$ or $\ket{H}_{-}$ next to $\ket{V}_{+}$. 
In this approach, 
one does not need to consider the time it takes for humans to process information; 
indeed, the polarization results are correlated in the data storage device 
whether experimentalists look at the data entry or not.
If an experimentalist observes the data results from a series of runs, 
he will conclude that $\ket{V}_{-}$ is always paired with $\ket{H}_{+}$ and $\ket{H}_{-}$ is always paired with $\ket{V}_{+}$.
If the above model is realized in nature 
and physicists believe that wavefunction collapse happens instantaneously, 
they will incorrectly conclude that causality is being violated in the Bell inequality experiments that have been conducted. 

The basic principle is that non-local instantaneous measurements are not possible 
given that the world is governed by special relativity.
One cannot simultaneously measure {\it and instantly compare} polarization measurements at $M_L$ and $M_R$
when $M_L$ and $M_R$ are space-like separated. 
A similar belief in the ability to perform instantaneous non-local measurements and a poor definition of distance 
has led to the incorrection conclusion\cite{SSExpandingUniverse}
that very distance points 
in an expanding universe can move apart faster than the speed of light.\cite{PeeblesCosmologyBook,Stuckey,Murdoch}

There is one strange thing about the above delayed collapse model: 
If the location of the right measuring device is moved farther to the right, 
then the time at which collapse can first begin (that is, the point $C$) moves further into the future. 
If there is no polarization measuring device on the right, 
then wavefunction collapse cannot happen even though a measurement 
has been made on the polarization of the left photon: 
The wavefunction must remain as in Eq.(\ref{OnePolarizationMeasurement}).
Hence, within this ``delayed'' model not all measurements can lead to wavefunction collapse. 

By the way, when one measures the photon polarization at $M_L$ in Figure \ref{fig:differentReferenceFrames}(a)
one cannot start wavefunction collapse in a region of a future light cone starting at $M_L$, meaning only with in $FLC_{M_L}$. 
This is because possible incompatibilities may arise 
when wavefunction collapse begins in a future light cone starting at $M_R$ (namely, within $FLC_{M_R}$) 
when the polarization of the right photon is measured: 
50\% of the time no problem arises because 
both measurements cause a collapse to $\Psi'_{HV}$ or to $\Psi'_{VH}$.
However, 50\% of the time one measurement causes a collapse to $\Psi'_{HV}$
while the other causes a collapse to $\Psi'_{VH}$, 
which leads a problem in making the wavefunctions ``match'' 
when they overlap in the future ``common causality'' region that starts at $C$, 
which is the intersection of $FLC_{M_L}$ with $FLC_{M_R}$.

Let us now discuss the case when unitarity and linearity 
rigorously hold in quantum field theory 
so that wavefunction collapse cannot happen.\footnote{The first physicist to seriously consider 
quantum mechanics without wavefunction collapse was H. Everett.\cite{EverettThesis}}
Then the final wavefunction for the experiments involving entangled photons 
is of the general form in Eq.(\ref{TwoPolarizationMeasurements}). 
Consider the case in which the measurement of the photon polarization 
produces photons that an experimentalist can see and for which two experimentalists are involved: 
one experimentalist observing the outcome on the left and the other observing the outcome on the right in 
Figure \ref{fig:EPRsetup} with $e^-$ and $e^+$ replaced by photons. 
Recall that in the measuring process, 
photons associated with the $\ket{H}$ polarization are emitted from a region that is spatially separated from 
the one associated with the $\ket{V}$ polarization. 
If the experimentalist sees the photons coming from the ``$\ket{H}$ region'' (respectively, ``$\ket{V}$ region''), 
then he concludes that the photon's polarization is $\ket{H}$ (respectively, $\ket{V}$). 
The photons from the ``$\ket{H}$ region'' interact with the quantum degrees of freedom inside the experimentalist 
differently from photons from the ``$\ket{V}$ region''. 
In this way, the wavefunction components involving the degrees of freedom 
within the experimentalist for the two cases end up being different. 
The fact that the experimentalist will think about what he has seen will produce additional differences between the two wavefunctions. 
This paragraph has introduced the wavefunction of the experimentalist (or his brain) in our analysis, 
and in the next paragraph we obtained a result concerning quantum mechanical limitations on human awareness. 
The issue of a connection between human thought and quantum mechanics 
has been raised before.\cite{QuantumMind,Stapp,JvonNeumann,Wigner,AlbertLoewer,Lockwood}

First, run the experiment for the case in which the initially state is only 
$\Psi_{HV} (\{ x_L \}, \{ x_R \}  , t) \ket{H}_{-} \ket{V}_{+}$.
The final state then becomes the first term in Eq.(\ref{TwoPolarizationMeasurements}), namely, 
$\Psi'_{HV} (\{ x'_L \}, \{ x'_R \} , t)$.
Since no wavefunction collapse is involved here, there is no issue about interpretation. 
The left experimentalist sees photons coming from the ``$\ket{H}$ region'' and is conscious of this, 
while the right experimentalist sees photons coming from the ``$\ket{V}$ region'' and also is conscious of this.
When they meet and report their results to each other, 
they both become conscious that the experimental result is for the situation 
in which the initial photon polarizations correspond to $\ket{H}_{-} \ket{V}_{+}$.
It is useful for future reference to define {\bf The Measurement Unawareness Result}: 
{\it The experimentalist with the quantum degrees of freedom in his body 
associated with one wavefunction component in the superposition
cannot be aware of another component in the superposition 
of the wavefunction governing a quantum measuring experiment.}
This statement is trivially true for the case just discussed 
in which the initial photon polarization is $\ket{H}_{-} \ket{V}_{+}$ because there is no superposition 
and hence no ``other component''. 
Next,  run the experiment for the case in which the initially state is 
$\Psi_{VH} (\{ x_L \}, \{ x_R \}  , t) \ket{V}_{-} \ket{H}_{+}$.
The final state then becomes the second term in Eq.(\ref{TwoPolarizationMeasurements}) or
$ - \Psi'_{VH} (\{ x'_L \}, \{ x'_R \} , t)$.
Following the same line of reasoning as in the first case, 
when they meet, they both become conscious that the initial photon polarizations correspond to $\ket{V}_{-} \ket{H}_{+}$.
As explained in the previous paragraph, 
the wavefunction involving the degrees of freedom of each experimentalist, however, is different for the two cases 
because photons coming from one direction produce a different interaction with the quantum degrees of freedom in a human body 
from photons coming from another direction. 
Basically, a wavefunction for a human that is aware that the experimental result is  $\ket{V}_{-} \ket{H}_{+}$  
cannot be the same as a wavefunction of a human that is aware that the experimental result is $\ket{H}_{-} \ket{V}_{+}$.
The Measurement Unawareness Result applies for this case too. 

When linearity strictly holds, 
the result for the entangled polarization case, 
in which the initial state is the left-hand side of Eq.(\ref{OnePolarizationMeasurement}), 
is the linear superposition of the above two cases not envolving entanglement. 
The form of the equation is that of Eq.(\ref{TwoPolarizationMeasurements}).
However, the first term in this equation involves a wavefunction for the two experimentalists 
that is different from the wavefunction for the two experimentalists in the second term. 
The two terms in Eq.(\ref{TwoPolarizationMeasurements}) are orthogonal  
because the two component wavefunctions are initially orthogonal 
due to the photon polarizations being orthogonal 
and remain so when quantum field theory is unitary 
because unitarity guarantees the constancy of inner products for all times (Eq.(\ref{Unitarity})). 
The situation is that of a generalized Schr\"odinger Cat involving humans. 
The wavefunction also correctly describes what happens in the Bell inequality experiments involving entangled photon polarizations: 
it represents a 50\% chance of having the experimentalists conscious that the initial photons polarizations correspond to $\ket{H}_{-} \ket{V}_{+}$ 
and a 50\% chance of having them conscious that the initial polarizations correspond to $\ket{V}_{-} \ket{H}_{+}$.
The question is: ``Is there anything wrong with this linear superposition of wavefunctions?''

First note that for the wavefunction component corresponding to the first term 
in which the initial photon polarization state is $\ket{H}_{-} \ket{V}_{+}$, 
an experimentalist (who is part by this wavefunction component) cannot be conscious of the existence 
of the second term (which corresponds to the initial state $\ket{V}_{-} \ket{H}_{+}$) because 
The Measurement Unwareness Statement holds for this situation.
Likewise,  for the wavefunction component corresponding to the second term 
in which the initial state is $\ket{V}_{-} \ket{H}_{+}$, 
a corresponding experimentalist cannot be conscious of the existence 
of the first term (which corresponds to the initial state $\ket{H}_{-} \ket{V}_{+}$).
Note that if an experimentalist were conscious of photons  {\it both} from the ``$\ket{H}$ region'' {\it and} the ``$\ket{V}$ region'',
then the structure of the wavefunction after the measurement but before the observation of the photons by the eye of an experimentalist
would involve the product $\Psi_{H \gamma's} \Psi_{V \gamma's}$, 
whereas the structure of the wavefunction that arises in the experiment is of the form $a \Psi_{H \gamma's} + b\Psi_{V \gamma's}$;
The former implies both photon emissions are present, while the latter implies that either one or the other emission occurs. 
Here, $\Psi_{H \gamma's}$ (respectively, $\Psi_{V \gamma's}$) represents a wavefunction for a set of photons  
originating from the ``$\ket{H}$ region'' (respectively, ``$\ket{V}$ region'').

A similar conclusion also holds if the results of the experiment are recorded, brought to a common location 
and then observed by a single experimentalist. 
Again one proceeds using the cases without entanglement and linearity:  
The evolution for the situation in which the initial polarization state is $\ket{H}_{-} \ket{V}_{+}$ is considered 
and then the evolution for the situation in which the initial state is $\ket{V}_{-} \ket{H}_{+}$ is treated.
It is reasonable to assume that the wavefunction of the experimentalist before observing the polarization result 
and the wavefunction of the rest of the degrees of freedom factorize 
so that the wavefunction of the experimentalist is initially not entangled.\footnote{Also see Eq.(\ref{factorization}) below.} 
However, when the experimentalist looks at the polarization result of a recording of the event, 
the light from the ``$\ket{H}_{-} \ket{V}_{+}$'' case enters his body and produces a different effect 
on the constituents in his brain 
and hence a different wavefunction for the degrees of freedom inside him 
from the case corresponding to the one for ``$\ket{V}_{-} \ket{H}_{+}$''. 
When a linear superposition of these two cases is constructed,
the situation is as above except that only one experimentalist is involved: 
The final result is a wavefunction that is a linear superposition of two components:
one involving a wavefunction which embodies the experimentalist believing the original polarizations are $\ket{H}_{-} \ket{V}_{+}$ 
and a second one involving a wavefunction which embodies the experimentalist believing the original polarizations are $\ket{V}_{-} \ket{H}_{+}$. 
In both cases, a component wavefunction of the experimentalist cannot convey to him the existence of the other component wavefunction
because The Measurement Unawareness Result holds. 
Again, the wavefunction correctly describes what happens experimentally: 
a 50\% chance of having the experimentalist conscious that the initial photon polarizations are $\ket{H}_{-} \ket{V}_{+}$ 
and a 50\% chance of having him conscious that the initial polarizations are $\ket{V}_{-} \ket{H}_{+}$. 
In the Many Worlds approach to quantum mechanics,\cite{ManyWorlds}
the importance of including the experimentalist in the analysis\cite{EverettThesis}  
and the fact that his wavefunction dependence is different depending on the outcome of an experiment 
is an accepted result.\cite{ManyWorlds,Tegmark}

Let us now compare the wavefunction collapse and the unitary quantum field theory\footnote{Unitary quantum field theory
(and unitary quantum mechanics) is defined by the existence of a unitary operator such as in Eq.(\ref{UnitaryOperator}), 
(often represented by a functional integral involving the Lagrangian) that governs the temporal evolution of states. 
However, most of the discussion in this section only uses the orthogonality preserving properties in Eq.(\ref{Unitarity}). 
In addition, it is assumed that $|\Psi(\{x\}, t)|^2$ provides the probability distribution in the $\{x\}$ 
and that $| {\langle  \Phi \vert   \Psi(t) \rangle }|^2$ is the probability of finding the system in the state $\Phi$ at time $t$.
}
approaches.
We have argued that if wavefunction collapse happens then it can only occur during events 
in which particles are produced or destroyed. 
Part of the reason why we are unable to rule out wavefunction collapse during these 
second-quantized processes is that we do not know the details of what transpires when a particle is created or destroyed. 
If wavefunction collapse can occur then there must be a non-perturbative violation of unitarity 
in quantum field theory and such a violation remains a speculation 
unless there is non-perturbative calculation to demonstrate it.
If wavefunction collapse can happen then one needs to understand why the Bohr rule holds,
namely why 
\begin{equation}
\begin{aligned}
a \Psi_1 + b \Psi_2 &\to \Psi_1 \ with \ probability \ { {|a|^2 } \over {|a|^2  + |b|^2 } }  \\
                     \ or &\to  \Psi_2 \ with \ probability \ { {|b|^2 } \over {|a|^2  + |b|^2 } }
 \ ,
\label{BohrRule}
\end{aligned}
\end{equation}
(where $\Psi_1$ and $\Psi_2$ are two normalized orthogonal states) can result when the wavefunction $a \Psi_1 + b \Psi_2$ collapses, 
and why some other formula does not hold, an example being  
$$
\begin{aligned}
a \Psi_1 + b \Psi_2 &\to \Psi_1 \ with \ probability \ { {|a|} \over {|a| + |b|} }  \\
                      \ or &\to \Psi_2 \ with \ probability \ { {|b|} \over {|a| + |b|} } 
\, .
\end{aligned}
$$
If wavefunction collapse happens instantaneously or almost instantaneously, 
then common sense locality and causality, which are fundamental principles of quantum field theory, 
appear to be violated 
and this leads to the EPR paradox. 
If one tries to preserve these two principles -- as we have attempted to do so in the delayed collapse model discussed above --  
then one is temporarily in the unitary case for which generalized Schr\"odinger Cats can exist. 
One of the ``advantages'' of instantaneous wavefunction collapse is that it eliminates generalized Schr\"odinger Cats; 
however, this ``advantage'' is absent at least temporarily in the above delayed wavefunction collapse model,
and, in certain circumstances, collapse cannot even happen as we have pointed out. 
More importantly, the main motivation for wavefunction collapse is the connection between the wavefunction and probability 
and the sense that humans have that there is only one macroscopic reality. 
If a measurement is taken, but wavefunction collapse does not immediately happen 
then what happens in the interim with the sense on behalf of an experimentalist that a definitive observation of the experimental outcome   
needs to eliminate the uncertainty associated with a wavefunction that is a superposition of the form $a \Psi_1 + b \Psi_2$? 
It is evident that a number of issues need to be overcome if wavefunction collapse is to take place in quantum field theory.

The results of the Bell inequality experiments are correctly produced by unitary quantum field theory. 
Furthermore, there is no EPR paradox in this case: 
Entangled polarization and spin states are oppositely correlated 
and a measurement of one does not cause an effect on the state of a distant polarization or spin. 
However, one consequence of unitary quantum field theory is 
the existence of generalized Schr\"odinger Cats. 
However, given that humans cannot be conscious of such superpositions, 
no ``observational'' contradiction arises. 
Given all the issues mentioned in the previous paragraph for the approach of wavefunction collapse, 
the solution using unitary quantum field theory is favored. 

If unitary quantum field theory is in agreement with experiment and functions fine without invoking wavefunction collapse, 
then why do so many physicists\cite{refDebate,Schlosshaueretal} believe in wavefunction collapse?
One possibility is human psychology. 
It is hard for us to believe that the wavefunction for a cat (or a human) can be in an ``extreme'' linear superposition 
when one has the overwhelming feeling that there is only one reality, that is, only one wavefunction component. 
However, if the structure of the wavefunction is a superposition 
in which each component wavefunction involves the quantum degrees of freedom in a human 
for which the human cannot be conscious of the other components (The Measurement Unawareness Result), 
then this is in agreement with this ``sense of uniqueness'' and the feeling that there is only one ``reality''.

To underscore the role of human thought, 
consider the possibility that the experiment is entirely carried out by machines. 
Humans could set it up, automate it and never look at the results. 
In this ``gedanken'' experiment  for the entangled photon-polarization system,  
one machine records the result ($\ket{H}_-$ or $\ket{V}_-$)
of the left photon-polarization-measuring device, 
another machine records the result ($\ket{H}_+$ or $\ket{V}_+$) of the right measuring device, 
the information is sent to a common location and the two results entered side-by-side in a data base. 
In unitary quantum field theory, 
the result is Eq.(\ref{TwoPolarizationMeasurements}), 
where the linear superposition involves two very different wavefunctions
for the degrees of freedom of the machines involved in the experiment.
The wavefunction in  Eq.(\ref{TwoPolarizationMeasurements}) reflects the result that the polarizations must be anti-correlated 
and that there are two possible outcomes each happening with a 50\% probability.
There is nothing intrinsically wrong with this wavefunction for this situation. 

In the previous paragraph, the machines had ``perception'' 
in that they could sense the polarizations and record the results.
Now assume that artificial intelligence has advanced to the stage 
in which the machines not only can record and compare polarizations,  
but they can also be made {\it aware} of the results. 
Then the wavefunction is still given as in Eq.(\ref{TwoPolarizationMeasurements}) 
but the wavefunction component associated with the first term in the linear superposition 
does not allow the machines to be aware of the existence of the second term in the linear superposition;  
a similar statement holds for the wavefunction associated with the second term in the linear superposition.
The difference between this case and the one of the previous paragraph 
is that the machines are ``conscious'' of the result 
and within each component the outcome of the experiment seems to be certain (as is the case in classical mechanics) 
even though a wavefunction of the form in Eq.(\ref{TwoPolarizationMeasurements}) involves uncertainty.

Next, the machines are taught some logic 
and the connection between probability and the absolute square of the wavefunction 
in quantum mechanics but not about the unitary nature of quantum mechanics. 
The machines are also made aware that the initial state is proportional to 
${( \ket{H}_- \ket{V}_+ -  \ket{V}_- \ket{H}_+ )/\sqrt{2}}$. 
The machines in the first component of Eq.(\ref{TwoPolarizationMeasurements}) 
being aware that the outcome of the experiment is $\ket{H}_- \ket{V}_+$ 
and ignorant of the existence of the second component of the linear superposition 
will (incorrectly) conclude that the outcome is certain 
and therefore the wavefunction must have collapsed to the first component 
at some point in order to maintain the connection between probability and the absolute square of the wavefunction. 
The above statements about awareness (or lack thereof) is contained in the wavefunction
of the degrees of freedom of the machines of the first component 
of the linear superposition of Eq.(\ref{TwoPolarizationMeasurements}). 
The machines in the second component of Eq.(\ref{TwoPolarizationMeasurements}) 
will come to an analogous conclusion, namely, 
that the wavefunction had to have collapsed to the second component 
(the second possibility in Eq.(\ref{CoordinatedWavefunctionCollapse})).
So when unitary quantum field theory is valid 
and machines with the intellectual ability of the first sentence of this paragraph are used 
to conduct the experiment, 
the result is a linear superposition of wavefunctions, 
in which each component wavefunction embodies the concept by the machines of wavefunction collapse. 

Finally, if the machines of the previous paragraph are additionally taught that unitarity {\it must} hold in quantum mechanics 
then logical thinking will eventually lead them to conclude 
that the wavefunction is as in Eq.(\ref{TwoPolarizationMeasurements}),
that the reason for the belief in wavefunction collapse in the previous paragraph 
is due to the inability to be aware of the other component in the linear superposition, 
and that a generalized Schr\"odinger Cat is occurring. 
When the machines associated with the first component of the superposition  
observe the output of a run to be $\ket{H}_- \ket{V}_+$, 
they will deduce 
that there must exist a linear superposition in which the another wavefunction component (the ``$\ket{V}_- \ket{H}_+$'' one) 
must be present in the ``full'' wavefunction even though they cannot be aware of it.
A similar statement holds for the machines associated with second component.

The above discussion involves what we call {\bf The Empirical Consciousness Statement}:
an experimentalist has the ability not only to observe a result but also to be able to be {\it aware} that he has observed the result. 
The Empirical Consciousness Statement is an empirical fact; it does not obviously follow from physics principles. 
Humans simply know they have this ability; why they have it is currently a mystery. 
In the above experiment involving entangled polarizations,
the experimentalist not only observes the outcome 
(either ``$\ket{H}_- \ket{V}_+$'' or  ``$\ket{V}_- \ket{H}_+$'') 
but also {\it knows} that he observed it.
Suppose we consider the first component of the superposition, 
that is, the one associated with the outcome ``$\ket{H}_- \ket{V}_+$''. 
If unitary quantum field theory holds and the experimentalist believes this, 
then he deduces that there is a linear superposition of two states 
as in Eq.(\ref{TwoPolarizationMeasurements}).
However, he also knows that 
the first-component wavefunction of the superposition permits him to be aware that it is the first component. 
Henceforth, the other component can have no effect on the quantum evolution of the first-component wavefunction 
due to unitarity (linearity and orthogonality). 
So the experimentalist can perform calculations 
{\it assuming} that the wavefunction had collapsed to the ``$\ket{H}_- \ket{V}_+$'' component
to determine the ``physics'' in his future.
A similar statement can be made for the second component of the superposition,  
in which the experimentalist observes the outcome to be ``$\ket{V}_- \ket{H}_+$''.\footnote{This is the way 
in which quantum mechanics incorporates conditional probabilities 
for the entangled photon Bell inequality experiments.}
In this sense, the use of wavefunction collapse can be regarded as a calculational convenience 
for a physicist involved in the experiment; 
a machine with artificial intelligence but without the ability to be aware of its observations 
would not be able to take advantage of this calculational simplification. 
It {\it is still incorrect} for the physicist to insist on wavefunction collapse -- that would lead to 
the previously discussed violations of properties and features 
of quantum mechanics and quantum field theory 
and generate the EPR paradox and other theoretical quandaries. 

The content of the last paragraph might remind one of the von Neumann-Wigner interpretation\cite{JvonNeumann,Wigner}
of quantum mechanics in which consciousness causes wavefunction collapse. 
However, in unitary quantum field, wavefunction collapse cannot happen. 
The Measurement Unawareness Result and The Empirical Consciousness Statement provide a basis in which human thinking tends 
to make one believe in wavefunction collapse. 
In this sense, consciousness is playing a role. 

Unitary quantum field theory provides an ``observational'' solution to the Measurement Problem,\cite{MeasurementProblem} 
which can be defined as the answers to two questions: 
(i) Does wavefunction collapse happen? 
(ii) If wavefunction collapse does not happen, why does it seem that the outcome of a quantum experiment 
yields a single result with certainty? 
Measuring devices appear to be governed by ``classical'' mechanics and devoid of the uncertainties of quantum mechanics 
that are present in the objects that they are trying to measure. 
Unitary quantum mechanics and quantum field theory say the answer to (i) is ``No'': wavefunction collapse does not happen. 
They also explain (ii):  
Consider a device that is capable of measuring $M$ different states $s_i$, $i = 1, \dots M$, 
where the $s_i$ are mutually orthogonal. 
We have described the situation in Section \ref{generalmeasurement}. 
When the initial state is a linear combination as in Eq.(\ref{LinearCombinationInitialState}) 
and an experimentalist observes the result, 
the final wavefunction in Eq.(\ref{LinearPicture}) is replaced by 
\begin{equation}
 \sum_j a_i s_i ( \{ y \}, t_f | S_i ) 
\, ,
\label{LinearPictureWithHuman}
\end{equation}
where $ \{ y \}$ includes all degrees of freedom of the system 
including those associated with the microscopic entities inside the experimentalist. 
Arguments based on linearity then lead to The Measurement Unawareness Result for each term, 
and Eq.(\ref{LinearPictureWithHuman}) correctly describes the experimental situation: 
there are $M$ possible outcomes for which the signal $S_i$ occurs with probability $| a_i|^2$ 
and in which the wavefunction for this outcome $s_i ( \{ y \}, t_f | S_i )$ 
does not permit the experimentalist associated with this wavefunction 
to be aware of any of the other terms in the superposition 
but The Empirical Consciousness Statement allows him to be aware of the one term to which he belongs.
This then leads to the appearance that the measuring device is producing a single certain result. 

It is important in achieving a viable understanding of quantum mechanics, 
for all degrees of freedom including those of humans to be incorporated properly into the analysis 
especially in avoiding conceptual difficulties with generalized Schr\"odinger Cats. 
Consider a Gerlach-Stern measurement of a spin-${1 \over 2}$ object 
in which the lower half of the detection screen -- if struck -- allows the release of a poison 
in the presence of a cat. 
If the spin of the spin-${1 \over 2}$ object is oriented in the $x$-$y$ plane, 
then there is a 50\% chance that the cat will die. 
Insisting on unitarity, the final wavefunction is of the form
\begin{equation}
a_{\uparrow} \Psi'_{\uparrow} (\{ x_c \}, \{ x_e \}) + a_{\downarrow} \Psi'_{\downarrow} (\{ x_c \}, \{ x_e \})  
 \ ,
\label{SchrodingerCatWavefunction} 
\end{equation}
where $|a_{\uparrow}|^2 = |a_{\downarrow}|^2 = 1/2$ and
where $\{ x_c \}$ are the quantum-mechanical degrees of freedom for all the microscopic consistuents of the cat 
and  $\{ x_e \}$ are the other relevant quantum-mechanical degrees of freedom 
for the neighboring environment, the poison and the experimental apparatus.\footnote{Actually, 
the variable sets $\{ x_c \}, \{ x_e \}$ might be different for the two components. 
For simplicity, we neglect this since it does not affect the argument.}
Here, $\Psi'_{\uparrow}$ (respectively, $\Psi'_{\downarrow}$) involves a wavefunction associated with a live (respectively, dead) cat. 
The wavefunction in Eq.(\ref{SchrodingerCatWavefunction}) correctly describes the situation. 
Now suppose that a human observes the experiment. 
It is incorrect to assume that the human is a ``by-stander'' and that the wavefunction factorizes 
\begin{equation}
\Psi_h (\{ x_h \}) (a_{\uparrow} \Psi'_{\uparrow} (\{ x_c \}, \{ x _e \}) + a_{\downarrow} \Psi'_{\downarrow} (\{ x_c \}, \{ x_e \})  )
 \ ,
\label{IncorrectSchrodingerCatHumanWavefunction} 
\end{equation}
where $\Psi_h$ is the wavefunction of the human and $\{ x_h \}$ are the microscopic degrees of freedom inside him. 
The problem with Eq.(\ref{IncorrectSchrodingerCatHumanWavefunction}) is that, 
when expanded, 
$\Psi_h (\{ x_h \}) \Psi'_{\uparrow} (\{ x_c \}, \{ x_e \})$ 
must involve a human wavefunction $\Psi_h$ in which the human has observed the cat to be alive 
and 
$\Psi_h (\{ x_h \}) \Psi'_{\downarrow} (\{ x_c \}, \{ x_e \})$ 
must involve a human wavefunction $\Psi_h$ in which the human has observed the cat to be dead. 
A single $\Psi_h$ cannot satisfy these two requirements. 
It is this type of reasoning that can lead to the belief that Schr\"odinger Cats cannot exist 
because it would require a human to observe a cat to be both dead and alive. 
When light from the event passes through the eyes of the human and 
interacts with his biological quantum-mechanical degrees of freedom, 
a difference wavefunction for the human arises depending on whether the cat is dead or alive.\footnote{The basic rule is 
that when something interacts with a Schr\"odinger Cat superposition, 
that thing becomes part of the Schr\"odinger Cat superposition. 
It does not matter whether the thing is microscopic or macroscopic 
or whether it is alive or inanimate, 
and the interaction can be direct or indirect.
If an experimentalist is unaware of his interaction with a generalized Schr\"odinger Cat superposition
then he is unable to take advantage of the wavefunction collapse simplification in determining his future evolution.} 
Unitary quantum field theory leads to a wavefunction of the form 
\begin{equation}
 a_{\uparrow} \Psi'_{\uparrow} (\{ x_h \}, \{ x_c \}, \{ x_e \}) + a_{\downarrow} \Psi'_{\downarrow} (\{ x_h \}, \{ x_c \}, \{ x_e \}) 
 \ .
\label{SchrodingerCatHumanWavefunction} 
\end{equation}
Now, the wavefunction correctly describes the situation:  
a 50\% chance of a live cat that is observed to be alive by a human (associated with $\Psi'_{\uparrow} (\{ x_h \}, \{ x_c \}, \{ x_e \})$)
and a 50\% chance of a dead cat that is observed to be dead by a human (associated with $\Psi'_{\downarrow} (\{ x_h \}, \{ x_c \}, \{ x_e \})$).
The Measurement Unawareness Result then guarantees that 
the human associated with either component of the superposition is only aware of ``his component''.
To emphasize the role of consciousness, 
suppose that the human who is present in the ``Schr\"odinger Cat event'' 
that leads to Eq.(\ref{SchrodingerCatWavefunction}) never observes or interacts with the cat. 
Then Eq.(\ref{IncorrectSchrodingerCatHumanWavefunction}) {\it does} correctly 
describe the situation: 
a 50\% chance that the cat is dead, a 50\% chance that it is alive, 
and a human who is unaware of (and unaffected by) the outcome. 

When a human experimentalist does not observe or interact with the degrees of freedom of an experimental outcome, 
then the functional dependence of the wavefunction  $\Psi$ on the degrees of freedom $\{ x_h\}$ of the human 
should be the same independent of the values of the degrees of freedom $\{ x_e\}$ associated with the experimental outcome. 
This means that ratio $\Psi (\{ x_h\}, \{ x_o\},  \{ x_e\})/ \Psi (\{ x_h\}, \{ x_o\},  \{ x'_e\})$, 
where $\{ x_e\}$ and $\{ x'_e\}$ correspond to two different sets of values of experimental outcome variables, 
should not have dependence on the $\{ x_h\}$. 
Hence,  $\Psi (\{ x_h\}, \{ x_o\},  \{ x_e\})/ \Psi (\{ x_h\}, \{ x_o\},  \{ x'_e\}) = f( \{ x_o\},  \{ x_e\},  \{ x'_e\})$ for some function $f$.
Here, $\{ x_o\}$ represent all the other relevant variables in the system and excludes $\{ x_h\}$ and $\{ x_e\}$.
Pick and fix a particular set of values for $ \{ x'_e\}$ and let  $\Psi_H (\{ x_h\}, \{ x_o\}) =  \Psi (\{ x_h\}, \{ x_o\},  \{ x'_e\} )$ 
and  $\Psi_E ( \{ x_o\},  \{ x_e\}) =  f( \{ x_o\},  \{ x_e\},  \{ x'_e\})$.
Then 
\begin{equation}
 \Psi (\{ x_h\}, \{ x_o\},  \{ x_e\}) = \Psi_H (\{ x_h\}, \{ x_o\}) \Psi_E ( \{ x_o\},  \{ x_e\}) 
\ , 
\label{factorization} 
\end{equation}
and the wavefunction factorizes. 
If $\Psi_E (\{ x_e\}, \{ x_o\})$ in Eq.(\ref{factorization}) involves a linear superposition of terms each corresponding to a different experimental outcome, 
then the wavefunction for the experimentalist is common to all the terms and 
the experimentalist cannot be aware of the experimental result.

Now return to the situation of the classical Schr\"odinger cat of Eq.(\ref{SchrodingerCatWavefunction}) 
and consider when two humans are present: 
one who observes the cat after the spin-${1 \over 2}$ object has hit the screen 
and one who does not observe it.
Use A as a label for the former human and use B for the latter.\footnote{An argument 
similar to the one in this paragraph 
avoids the difficulties with Wigner's Friend\cite{WignersFriend} 
as H. Everett pointed out.\cite{EverettThesis} 
Here, Wigner is spectator B and his friend is observer A.
}  
Then the final form of the wavefunction is 
\begin{equation}
 \Psi_B (\{ x_B \}, \{ x_o\} )
(        a_{\uparrow} \Psi'_{\uparrow} (\{ x_A \}, \{ x_c \}, \{ x_o\}, \{ x_e \}) + 
 a_{\downarrow} \Psi'_{\downarrow} (\{ x_A \}, \{ x_c \}, \{ x_o\}, \{ x_e \}) )
 \ ,
\label{SchrodingerCatTwoHumanWavefunction} 
\end{equation}
which is Eq.(\ref{SchrodingerCatHumanWavefunction}) with the addition of a spectator (human B) 
and now includes the effect of common degrees of freedom $\{ x_o\}$ 
not involving $\{ x_A \}$, $\{ x_B \}$, $ \{ x_c \}$ and $ \{ x_e \}$.
As discussed previously, observer A is aware of what he sees and can invoke wavefunction collapse 
as a calculational method. 
Spectator B cannot use wavefunction collapse 
and this situation will persist unless observer A informs B of the result, 
in which case the effect of this information on B will change the nature of the wavefunction 
associated with the degrees of freedom inside B differently for the two terms in Eq.(\ref{SchrodingerCatTwoHumanWavefunction}) 
leading to 
\begin{equation}
        (a_{\uparrow} \Psi''_{\uparrow} (\{ x_A \}, \{ x_B \}, \{ x_c \}, \{ x_o\}, \{ x_e \}) + 
 a_{\downarrow} \Psi''_{\downarrow} (\{ x_A \}, \{ x_B \}, \{ x_c \}, \{ x_o\}, \{ x_e \}) )
 \ .
\label{SchrodingerCatTwoKnowingHumanWavefunction} 
\end{equation}
Now due to The Empirical Consciousness Statement, 
spectator B will know which of the two components he belongs to, 
that is, he will know whether the cat is dead or not, 
and he too can invoke wavefunction collapse as a calculational tool 
in ``computing the future evolution of his wavefunction component''.

Additionally, an experimentalist might naively ask, 
``If the degrees of freedom of my body are associated with a linear superposition of components,
then how do I know which component corresponds to me?'' 
This is a natural question to ask because humans have self awareness. 
Because of The Empirical Consciousness Statement, 
the answer to the question is that the degrees of freedom of each component wavefunction permit 
the experimentalist associated with the $i$th component in Eq.(\ref{LinearPictureWithHuman}) 
to be aware that he is associated with the $i$th component. 
The experimentalist might follow with the question, 
``If I know who I am, who is associated with a component that does not correspond to me, that is, 
a component $j$ with $j \ne i$.
The answer is that soon after the measurement is made, 
it is extremely likely that there is a living human associated with the $j$th component 
and that that living person is almost ``the same'' as the experimentalist -- 
what one might call a ``Schr\"odinger Twin''.
However, given enough time it is likely that the ``twin'' will become manifestly different 
including the likelihood that the quantum constituents inside the two bodies no longer perfectly coincide 
because of second-quantized particle creation and destruction events. 
In Appendix B, 
we present some more examples of the structure of wavefunctions associated with 
quantum experiments of binary outcomes under the assumption that unitary quantum field theory is correct. 

Our discussion of unitary quantum field overlaps considerably with the Many Worlds approach to quantum mechanics\cite{ManyWorlds} 
and with the Many Minds interpretation.\cite{AlbertLoewer}
Common to these is that wavefunction collapse does not happen;\cite{EverettThesis}
the Many Worlds and Many Minds approaches also assume unitarity-respecting evolution of quantum mechanics 
and emphasizes the importance of including observers (e.g. humans) into the wavefunction when experiments are conducted.
However, assigning ``worlds'' (or ``minds'') to linear superpositions of wavefunctions 
and assuming that certain events cause ``branching'' into these worlds 
may be more confusing than helpful. 
If it is assumed that a situation such as the one described by Eq.(\ref{DistinctiveQuantumStates}) is supposed 
to correspond to a branching of the world at $t_0$ into two branches $A$ and $B$ 
then the probabilities for the branching processes must be $|\Psi_A (\{x_A\}, t)|^2$ and $|\Psi_B (\{x_B\}, t)|^2$ respectively. 
However, this Born rule must be postulated.\cite{ManyWorldsBornRule} 
The wavefunction in Eq.(\ref{DistinctiveQuantumStates}) has no issues with the Born rule. 
For the entangled spin-${1 \over 2}$ system in which the initial state is an eigenstate of angular momentum, 
another issue arises: 
In each ``world'', angular momentum would be violated 
because one world would be associated with the $ \ket{\uparrow}_{-} \ket{\downarrow}_{+}$ state component 
while the other  one would be associated with the $ \ket{\downarrow}_{-} \ket{\uparrow}_{+}$ one. 
It is necessary to keep the two worlds ``together'' to maintain angular momentum conservation.
When too broad a definition of ``world'' is used, 
the Many Worlds approach runs into difficulties as to which basis\cite{BasisProblem} 
one should use for the linear superposition: 
It is possible to express a wavefunction in terms of a linear combination 
of mutual orthogonal wavefunction components in many ways. 
Hence, the ``worlds'' in the Many Worlds approach depends on how one expands the wavefunction, 
which makes little sense since this is up to the whim of the physicist. 
These issues do not arise if one simple sticks to using the wavefunction $\Psi (\{x\}, t)$, 
where $\{x\}$ contains all the quantum degrees of freedom.
A given wavefunction $\Psi (\{x\}, t)$ for a closed system
describes the quantum mechanical situation unambiguously. 
If the closed system is the entire universe, 
$\Psi (\{x\}, t)$ is the wavefunction of that universe and not the wavefunction of many universes. 
It is not necessary to express $\Psi (\{x\}, t)$ as a linear superposition of terms 
if one does not want to. 
The probability distribution in the $\{x\}$ at time $t$ is given by $|\Psi (\{x\}, t)|^2$ 
and the probability that the system is in a state $\Phi_0$ at time $t$ is $|\bracket {\Phi_0}{\Psi (t)} |^2$, 
regardless of whether the wavefunction is written as a linear superposition or not.\footnote{We {\it assume} that 
the probability distribution is given by the absolute square of the wavefunction 
and do not try to derive it from some other principles 
as is sometimes attempted in the Many Worlds approach.\cite{ManyWorlds} 
} 
However, expressing the wavefunction as a linear superposition can be calculationally convenient 
when the evolution of individual components can be more easily computed 
or can be useful for gaining insight by a physicist into a particular problem.\footnote{Our use of linear superpositions in this section 
is an example -- the purpose being to show that unitary quantum field theory correctly describes quantum measurements,
and, in particular, the results of the experiments involving the EPR paradox.}  
Finally, it is important to incorporate The Empirical Consciousness Statement 
(and also The Measurement Unawareness Result) to obtain the ``observational'' solution of the Measurement Problem. 

Roughly speaking, the difference between Many Worlds quantum mechanics and unitary quantum mechanics 
when applied to an experimentalist performing a quantum experiment with the possibility of various outcomes 
is that the former involves one experimentalist and many worlds while the latter involves one world 
and many ``experimentalists'' or, more precisely, many configurations of the quantum constituents of the experimentalist.

It is interesting to return to the Gedanken experiment concerning the Heisenberg uncertainty principle 
for the entangled spin-${1 \over 2}$ system and analyze it from the point of view of unitary quantum mechanics and field theory. 
Recall this corresponds to the setup in Figure \ref{fig:setupMoving} 
when the left experimentalist measures the spin of the electron in the $z$-direction  
and the right experimentalist measures the spin of the positron in the $y$-direction. 
The form of the wavefunction after a measurement is\footnote{The constants in front of each wavefunction 
component of the superposition could be absorbed into the wavefunctions, 
but we express it in this way for pedagogical purposes.}
 \begin{equation}
 {1 \over 2} (- i \Psi'_{\uparrow \rightarrow} (\{ x' \} ) + i \Psi'_{\uparrow \leftarrow} (\{ x'\}) - 
   \Psi'_{\downarrow \rightarrow} (\{ x' \} )  -  \Psi'_{\downarrow \leftarrow} (\{ x' \} ) )  
 \ ,
\label{FinalFormEntangledSpinWavefunction} 
\end{equation}
where $\{ x' \}$ contain the quantum degrees of the electron, the positron, the two spin measuring devices, and the experimentalists, 
and $\uparrow$ (respectively, $\downarrow$) represents an up-spin (respectively, down-spin) measurement of the electron 
in the $z$-direction corresponding to the initial state $\ket{\uparrow}_{z -}$ (respectively, ($\ket{\downarrow}_{z -}$), 
and $\rightarrow$  (respectively, $\leftarrow$) represents an up-spin (respectively, down-spin) measurement of the positron 
in the $y$-direction corresponding to the initial state $\ket{\uparrow}_{y +}$ (respectively, $\ket{\downarrow}_{y +}$).
(The $-$ and $+$ subscripts still indicate electron and positron or particle moving to the left and particle moving to the right.)
Consider the first term in Eq.(\ref{FinalFormEntangledSpinWavefunction}). 
The wavefunction of the experimentalists contains their observations that the electron's spin is up in the $z$-direction 
and that the positron's spin is up in the $y$-direction. 
Since both measurements occurred first in their respective rest frames at a time 
when the spin content of the wavefunction was 
$(\ket{\uparrow}_{-} \ket{\downarrow}_{+} -  \ket{\downarrow}_{-} \ket{\uparrow}_{+})$, 
meaning that electron and positron spins are oppositely orientated independent of the axis of spin quantization, 
the wavefunction of the experimentalists contains the thought that the electron's spin was measured to be
both up in the $z$-direction and down in the $y$-direction 
(or that the positron's spin was measured to be both up in the $y$-direction and down in the $z$-direction) 
contradicting Heisenberg's uncertainty principle in Eq.(\ref{HeisenbergUncertaintyForSpinEq}) 
that two components of spin can be precisely measured. 
This ``thinking'' is a result of the two experimentalists not being able to be aware (due to The Measurement Unawareness Result) 
of the three other terms in Eq.(\ref{FinalFormEntangledSpinWavefunction}). 
Indeed, if wavefunction collapse were to occur, 
rending the final wavefunction as just the first terms in Eq.(\ref{FinalFormEntangledSpinWavefunction}) 
then this thinking would be correct and aspect (iii) of the EPR paradox would arise.
However, if the experimentalists believe in the unitarity of quantum mechanics then 
they can deduce the presence of the three other term in the wavefunction. 
The first two terms are compatible with the electron's spin being up and the positron's spin being down in the $z$-direction
since $\ket{\downarrow}_{z +} = (-i \ket{\uparrow}_{y +} + i \ket{\downarrow}_{y +})/\sqrt{2}$ 
while the first and third terms are compatible with the positron's spin being up and the electron's spin being down in the $y$-direction 
since $ - i \ket{\downarrow}_{y -} = (-i \ket{\uparrow}_{z -} - \ket{\downarrow}_{z -})/\sqrt{2}$.
With all four terms present, Eq.(\ref{FinalFormEntangledSpinWavefunction}) implies
$\Delta S_y = \Delta S_z = \hbar/2$ for the electron (with the $\Delta S_y = \hbar/2$ result
being deduced from the knowledge that the spin of the electron must be opposite to that of the positron)
and therefore Eq.(\ref{HeisenbergUncertaintyForSpinEq}) is automatically satisfied 
since the maximum value that $| \left\langle S_x \right\rangle |$ can be is $\hbar/2$. 
The same result holds for the positron: $\Delta S_y = \Delta S_z = \hbar/2$ and there is no 
violation of the Heisenberg uncertainty principle.
If ``Worlds'' are associated with each wavefunction component in the superposition of Eq.(\ref{FinalFormEntangledSpinWavefunction}), 
then the ``splitting'' into worlds created by the measurement of each experimentalist 
in that experimentalist's rest frame happens 
before the ``splitting'' into worlds created by the measurement of the other experimentalist 
for the setup in Figure \ref{fig:setupMoving}
and the same reasoning used in the context of wavefunction collapse works
to show that the Heisenberg uncertainty principle is violated in each of the four worlds.\footnote{By the way, 
the ``delayed'' wavefunction collapse model discussed earlier in this section avoids 
violating the Heisenberg uncertainty principle because one cannot use special relativity 
to create a ``timing'' effect with the measurements, that is, that each of the two measurements 
determined the spin in a definite direction and that both occurred before the other in their respective rest frames. } 
Hence, it is conceptually better not to try to assign a separate World to each term; 
the ``entire'' wavefunction by itself correctly describes the situation. 

Equation (\ref{FinalFormEntangledSpinWavefunction}) illustrates some of the results obtain in the first part of this paper:
(a) Quantum uncertainty at the microscopic level is transferred to uncertainty at the macroscopic level; 
The former, which is uncertainty in the spins as represented by the wavefunction 
${ 1\over 2} (-i \ket{\uparrow}_{z +} \ket{\uparrow}_{y +} + i \ket{\uparrow}_{z -} \ket{\downarrow}_{y +}
 -  \ket{\downarrow}_{z -}  \ket{\uparrow}_{y +}  -  \ket{\downarrow}_{z -}  \ket{\downarrow}_{y +} )$
becomes the latter, which corresponds to the four terms in the linear superposition
of the generalized Schr\"odinger Cat in Eq.(\ref{FinalFormEntangledSpinWavefunction}), 
each of which occurs with probability ${1 \over 4}$. 
(b) When one considers any one term in Eq.(\ref{FinalFormEntangledSpinWavefunction}), 
the signal in the macroscopic measuring devices indicating that the spins have a particular value 
($\ket{\uparrow}_{z +} \ket{\uparrow}_{y +}$ or $\ket{\uparrow}_{z -} \ket{\downarrow}_{y +}$, etc.)
does not necessarily mean that the spins have or had this value. 
(c) Multiple measurements are needed to determine the spin wavefunction. 
Only by performing the experiment many times can one determine that it 
consists of four terms whose coefficients have a magnitude of $1/2$. 
A different experimental setup is needed to determine the phases of these coefficients 
in a procedure analogous to the one used in Section \ref{SGmeasurement} to measure spin ${1 \over 2}$. 

The reader may have noticed that there is almost a one-to-one correspondence between 
issues (e.g. Born Rule, possible violation of angular moment and Heisenberg's uncertainty principle) 
with the Many Worlds approach and issues with wavefunction collapse. 
H. Everett originally argued that wavefunction collapse was inconsistent because of Wigner's friend.\cite{EverettThesis} 
The same argument applies to the Many Worlds approach: 
If Wigner's friend (human A) conducts an experiment inside a box then the world needs to branch for him 
but for Wigner (human B) the world should not branch.
In unitary quantum mechanics, there is no issue: 
The wavefunction in Eq.(\ref{SchrodingerCatTwoHumanWavefunction}) correctly describes the situation.
It is evident that same human B needs to be in both worlds, while a different human A appears in the two worlds. 
Even the EPR paradox re-emerges for the Many Worlds approach: 
In each world there is ``spooky'' action at a distance (EPR aspect (ii)). 
To avoid this, the ``worlds'' need to be kept together 
and this is accomplished in unitary quantum mechanics by using the full wavefunction. 
In the entangled spin-${1 \over 2}$ system, 
the wavefunction is a linear superposition of two components from ``start'' to ``finish''; 
there is no ``branching moment'', 
which is especially evident if more than one experimentalist is involved 
with experimentalists becoming aware of the outcome at different times. 

\section{Discussion and Conclusions}
\label{conclusion}

Sections \ref{AngularMomentumViolation} and \ref{UncertaintyPrincipleViolation} 
showed that the existence of a Spin-${1 \over 2}$ Collapse Measurement 
for the entangled spin-${1 \over 2}$ system can in idealized Gedanken experiments violate angular momentum, 
create cause-and-effect problems concerning the collapse of the wavefunction, 
and violate the Heisenberg uncertainty principle in first-quantized quantum mechanics. 
Given the known fact that it also violates unitarity, 
wavefunction collapse cannot occur for this system within pure first-quantized quantum mechanics. 

If wavefunction collapse cannot happen for the entangled spin-${1 \over 2}$ system, 
one should question the validity of the wavefunction collapse picture of measurement in general 
and whether such a collapse can occur in other measurements 
within first-quantized quantum mechanics. 
In any system with entangled, non-commuting observables, 
the use of moving measuring devices allows the violation of an Heisenberg uncertainty principle. 
This is true for position and momentum in the original EPR paper:\cite{epr}
Moving measuring devices allow a right-observer to measure momentum 
and a left-observer to measure position in such a way that both measurements take place first in the rest frames of each observer. 
A correlation in positions in the EPR setup then allows the right-observer to deduce the positon of the left object 
from the measurement of the right-observer. 
Then violation of the Heisenberg uncertainty relation $\Delta x\Delta p_x \ge \hbar /2$ for the left object occurs in this Gedanken experiment, 
suggesting that wavefunction collapse also cannot happen in this case. 

The issue of wavefunction collapse during a measuring process has been open to debate.\cite{refDebate,Schlosshaueretal}
Our arguments against wavefunction collapse in first-quantized quantum mechanics are therefore of value. 

If wavefunction collapse does not occur, 
then aspects (ii) (``spooky action at a distance'') and (iii) (possible violation of the Heisenberg uncertainty principle) 
of the EPR paradox are not problematic.
In addition, the Converse Collapse Statement then indicates that 
a single measurement does not necessarily definitively determine the nature of a quantum state, 
that in many (indeed most) cases, multiple measurements are needed to determine the it 
even when the signal of a device seems to indicate that it is in a particular state.
Furthermore, quantum-mechanical uncertainty at the microscopic scale 
can be (and usually is) transmitted to uncertainty at the macroscopic scale.
This is an important conclusion concerning 
how measurement\cite{quantummechanicsBookDirac,quantummechanicsBookBorn,JvonNeumann} works in quantum mechanics.

If wavefunction collapse does occur, 
then in most cases, multiple measurements are {\it still} needed to determine a quantum state 
but the reason is different than the one for unitary quantum mechanics:  
Wavefunction collapse for a single measuring event usually leads to loss of information about the state.

In Section \ref{SGmeasurement}, we illustrated the need for multiple measurements  
using a device to measure spin-${1 \over 2}$. 
We provided a criterion, which we call the Quantum Measurement Rule, 
at the end of Section \ref{generalmeasurement} 
that indicates when a single measurement suffices to determine the quantum state 
and when multiple measurements are needed:
If wavefunction collapse is needed to explain 
the ``certainty'' of an experimental event, 
then multiple measurements are required to determine the state. 
Although we obtained the Converse Collapse Statement and Quantum Measurement Rule 
within first-quantized quantum mechanics, 
the two also apply to unitary quantum field theory. 

The EPR paradox has not been tested in first-quantized quantum mechanics. 
All experiments to date involve second-quantized processes -- all with photons. 
This led us to consider the issue of wavefunction collapse in quantum field theory.
In Section \ref{SecondQuantization}, 
we examined whether the arguments used to rule out wavefunction collapse 
in first-quantized mechanics apply to quantum field theory. 
We were unable to demonstrate that wavefunction collapse cannot 
happen during processes in which particles are created or destroyed. 
So, if wavefunction collapse occurs, it most likely happens when there is a change in Fock-space particle content.\cite{Melkikh} 
As in first-quantized quantum mechanics, 
the existence of wavefunction collapse 
implies that unitarity in quantum field theory does not hold.
Unitarity is valid in quantum field theory to all orders in perturbation theory. 
Therefore, wavefunction collapse requires a non-perturbative violation of this property. 
The issue of locality and causality is more difficult; 
if wavefunction collapse happens instantaneously, 
then both are violated in quantum field theory and the EPR paradox emerges. 
However, it is possible to develop a ``delayed'' wavefunction collapse model 
in which locality and causality, which are basic principles of quantum field theory, 
are maintained. 
This delayed model of wavefunction collapse is actually in agreement with the Bell inequality experiments 
involving entangled particle spins and photon polarizations 
when one takes into account that comparing distant measurements of spins or polarizations 
cannot be conducted instantaneously but must be brought or communicated to a location 
that is within the future light cones of each measurement. 
However, in an entangled system sometimes it is necessary to indefinitely delay the wavefunction collapse 
associated with a measurement. 
The delayed model also has, at least temporarily, macroscopic generalized Schr\"odringer Cats. 
For these reasons, 
there is a certain awkwardness with this delayed model of wavefunction collapse.
This led us to reconsider the consequences of unitary quantum field theory 
in which wavefunction collapse cannot take place. 

Within unitary quantum field theory,
we found that when the wavefunction of the experimentalists and the measuring equipment 
are incorporated into the analysis, 
and we insist on linearity ``to the extreme'', 
a consistent picture of the Bell inequality and EPR experiments emerges. 
However, linearity led to the conclusion that 
although the wavefunction involves a linear superposition of component wavefunctions 
representing seemingly ``incompatible'' situations, 
an experimentalist associated with the wavefunction of one component of the superposition 
cannot be aware of any of the other components, 
a property that we call The Measurement Unawareness Result. 

In the above, the ability of an experimentalist to be aware of the outcome of an experiment 
plays an interesting role. 
By replacing human experimentalists by machines with varying abilities of artificial intelligence,
we arrived at an understanding of why some physicists believe in wavefunction collapse 
even if it does not occur in nature. 
Indeed, if one focuses on the connection between quantum mechanics and probability and downplays quantum-mechanical unitarity,
then it is ``natural'' to think that  wavefunction collapse is happening.
Even stranger is the role of human thought in resolving the Measurement Problem: 
it partially depends on humans not only being able to observe an experimental result 
but also being aware of this (The Empirical Consciousness Statement), 
something that is simply an observational truism. 
The upshot is to create the feeling that the world is classical 
because human senses are incapable of detecting microscopic quantum effects 
and also unable to be cognizant of certain quantum uncertainties that exist at the macroscopic scale.

A necessary consequence of unitary quantum field theory is the existence of generalized Schr\"odinger Cats;
indeed, one expects them to be commonplace.
However, their existence is consistent with human observation because of The Measurement Unawareness Result: 
humans cannot be conscious of them in experiments. 
Indeed, a wavefunction of linear superpositions of orthogonal components 
correctly represents the experimental situation. 
For example, 
in the case of the entangled spin-${1 \over 2}$ system discussed in the first part of this article, 
the wavefunction after the measurement becomes a linear superposition of two terms: 
The first wavefunction component of the superposition involves the degrees of freedom of experimentalists 
in which they are conscious of observing the measurement of the spin state to be $\ket{\uparrow}_- \ket{\downarrow}_+$  
but cannot be aware of the second wavefunction component, 
in which the degrees of freedom of experimentalists 
lead them to be conscious of observing the spin state to be $\ket{\downarrow}_- \ket{\uparrow}_+$. 
The inner product with itself of each component is ${1 \over 2}$ 
(indicating that each happens with 50\% probability). 
Furthermore, if, for example, the experimentalists are aware that the outcome is  $\ket{\uparrow}_- \ket{\downarrow}_+$, 
then they can invoke wavefunction collapse to $\ket{\uparrow}_- \ket{\downarrow}_+$ as a calculation tool but 
within unitary quantum field theory the existence of another wavefunction component 
in which the experimental outcome is $\ket{\downarrow}_- \ket{\uparrow}_+$ cannot be denied.
The Empirical Consciousness Statement plays a role here: 
If the outcome is recorded by a machine, then the measurement has taken place, 
but if the experimentalists never observe the result then 
they cannot make the wavefunction collapse ``simplification''.
In unitary quantum field theory with the above understanding, 
there is neither an EPR paradox nor a Measurement Problem. 

The resolution of the Measurement Problem in unitary quantum field theory 
is the result of linearity, which yields The Measurement Unawareness Result, 
and The Empirical Consciousness Statement. 
Since first-quantized quantum mechanics is also unitary, 
The Measurement Unawareness Result holds within this theory too. 
Given that The Empirical Consciousness Statement is a general observational fact, 
the Measurement Problem is also resolved in first-quantized quantum mechanics. 
Since unitary quantum field theory provides a satisfactory solution to the Measurement Problem,  
there is not reason to assume wavefunction collapse happens 
even through we have been unable to rule it out during particle creation and destruction events.

One arrives at a picture in which unitary quantum mechanics and quantum field theory 
are consistent with observations and without paradoxes. 
One of the surprising conclusions is that 
not only does quantum mechanical uncertainty 
extend to macroscopic scales but it involves 
distributions in the degrees of freedom of the wavefunction 
that can be macroscopically very disparate 
but humans are unable to be aware of these quantum mechanical effects. 

One may ask if there is any experimental evidence that unitary quantum field theory is correct 
and that wavefunction collapse is wrong. 
The interpretation of an experiment is often framed by current theoretical understanding. 
If unitary quantum field theory (or unitary quantum mechanics) had been the accepted theory from the beginning 
and someone had subsequently proposed the idea of wavefunction collapse,
then experiments such as the EPR ones involving entangled photons 
might have been proposed as a way of seeing if wavefunction collapse is problematic. 
The results of these EPR experiments under the belief that wavefunction collapse happens 
would have led  to the conundrum of ``spooky'' action at a distance 
and the appearance of loss of causality and locality, 
and many physicists would have thus concluded that wavefunction collapse is inconsistent with experiment. 
Instead, the historical path has been that wavefunction collapse (or something quite similar) 
happens during a measurement, 
and when this is the accepted view, then the EPR experiments generate the belief that something is wrong with
(or at least mysterious about) quantum mechanics.

\section*{Acknowledgments}

I thank Daniel Greenberger for early conversations on the part of this work involving the EPR paradox. 
I thank Machiel Blok and Ronald Hanson for discussions about their Bell inequality experiment involving entangled electron spins. 

\appendix
\section{A Device to Measure Spin-$1 \over 2$ that Is an Eigenstate of Angular Momentum}
Since there is no spin-$1 \over 2$ measuring device with $J=0$, 
the question arises as to the existence of such a device for the case $J>0$.
This is needed as part of the argument concerning violation of angular momentum in Section \ref{AngularMomentumViolation}. 
The only requirement is that when the incoming spin is up the device measures the spin to be up, 
and when the incoming spin is down, it measures it to be down.
As a gedanken experiment, 
one only needs  to require the measuring apparatus not to be in conflict with the laws of physics. 
This means there is considerable flexibility in its construction. 
Many such spin measuring devices are likely to exist. 
Here, we provide an example. 

The device consists of a large number of spin-$1 \over 2$ objects with all spins pointing up. 
If there are $N_s$ such spins, 
then it is an eigenstate of angular momentum corresponding to $\ket{ {{N_s} \over 2} ,  {{N_s} \over 2} }$. 
We assume that the device's objects only interact with electrons and this occurs through spin exchange: 
$\ket{\downarrow}_e \ket{\uparrow}_s \to \ket{\uparrow}_e \ket{\downarrow}_s$ 
where $e$ indicates ``electron'' and $s$ is any one of the $N_s$ spins of the device. 
We assume that the device has sufficiently many spin-$1 \over 2$ objects that a down-spin electron 
passing through it will flip its spin with virtually 100\% probability. 
After the electron passes through the device,
the electron's spin will always be up. 
The existence of a single down-spin among the $N_s$ spin-$1 \over 2$ objects of the device
indicates that the spin of the electron was originally down. 
If all spins in the measuring device point up, 
then the spin of the electron has been measured to be up. 

Instead of using spin exchange, one can use charge. 
Let the objects of the above device be neutral.
If an up electron enters the device then it replaces one of the neutral spin-$1 \over 2$ up objects and ``passes'' it momentum to it.
If a down electron enters the device then it simply travels through it. 
If after the measurement, the device has charge of $-e$, then the spin of the electron was up;
If the device remains neutral, then the electron's spin was down.

For the situation in Figure \ref{fig:EPRsetup},
one cannot guarantee that the electron is emitted in a particular direction. 
Hence, one needs to have $4 \pi$ solid angular coverage. 
This requires a spherical shell of spin-$1 \over 2$ objects of a certain thickness 
at a certain distance from the source. 

If one also wants to measure the positron's spin, 
a second set of spin-$1 \over 2$ objects that only interact with positrons can be used.
It can be located at a second spherical shell that does not overlap with the one used to measure the spin of an electron. 
If there are $N_s'$ spins involved in measuring the positron's spin, 
then the initial state it an eigenstate of angular momentum corresponding to 
$\ket{ {{({N_s} + {N_s'})} \over 2} , {{({N_s} + {N_s'})} \over 2} }$.

\section{Further Examples of Unitarity-Respecting Wavefunctions}
In this section we provide three additional examples of how the Measurement Problem is resolved 
in unitary quantum field theory and quantum mechanics. 
We consider the case of a binary outcome 
and, for simplicity, the measurement of the spin in the $z$-direction of a spin-$1 \over 2$ object, 
although the results apply equally to the entangled spin-$1 \over 2$ system\footnote{See discussion 
in the third paragraph before the last paragraph in Section \ref{conclusion}.}
considered in the first part of this paper 
and the entangled photon-polarization case of Section \ref{SecondQuantization}.

Consider the situation in the paragraph containing 
Eqs.(\ref{SchrodingerCatTwoHumanWavefunction}) and (\ref{SchrodingerCatTwoKnowingHumanWavefunction}) 
in which one experimentalist observes whether a spin-${1 \over 2}$ object is up or down in a measurement 
but in which no cat is involved or potentially poisoned. 
As before, the experimentalist reports the result of the outcome to a spectator (human B) 
who is far away from the experiment. 
Unbeknownst to the spectator, the experimentalist is always going to {\it say} that the spin is up 
even if he measures it to be down. 
The experimentalist prepares an email message beforehand that reads, ``The spin was measured to be up.'' 
The wavefunction initially can be written as 
\begin{equation}
 \Psi_B (\{ x_B \}, \{ x_o\} )  \Psi_A (\{ x_A \}, \{ x_o\}, \{ x_e \} ) 
  (a_{\uparrow}  \ket{\uparrow} +  a_{\downarrow} \ket{\downarrow} )
 \ .
\label{InitialTwoPersonWavefunction} 
\end{equation}
We include the non-spin degrees of freedom of the spin-${1 \over 2}$ object 
(and those of the experiment apparatus) in $ \{ x_e \}$. 
Here, $\{ x_A \}$ and $\{ x_B \}$ are the quantum degrees of freedom inside the experimentalist and the spectator respectively, 
and $ \{ x_o\}$ are other degrees of freedom. 
After the measurement is made and the email message sent to the spectator, 
the structure of the wavefunction is similar to Eq.(\ref{SchrodingerCatTwoHumanWavefunction}) 
(but without the cat): 
\begin{equation}
 \Psi'_{B\uparrow} (\{ x_B \}, \{ x_o\} )
(        a_{\uparrow} \Psi'_{\uparrow} (\{ x_A \}, \{ x_o\}, \{ x_e \}) + 
 a_{\downarrow} \Psi'_{\downarrow} (\{ x_A \}, \{ x_o\}, \{ x_e \}) )
 \ , 
\label{FinalTwoPersonWavefunction} 
\end{equation}
where $\Psi'_{B\uparrow} (\{ x_B \}, \{ x_o\} )$ is the wavefunction for the human spectator that would result 
if the initial spin state were purely up ($ a_{\uparrow} = 1$, $a_{\downarrow} = 0$).
Even though the spectator was ``informed'' of an outcome of the experiment, 
his wavefunction still factorizes. 
Furthermore, if he is convinced that unitary quantum field theory is correct, 
he will incorrectly conclude that there is another term in wavefunction in Eq.(\ref{FinalTwoPersonWavefunction}) 
that he cannot be aware of but in which the thoughts associated with the quantum degrees of freedom inside his body 
indicate that the spin is down. 
In addition, he will also incorrectly conclude that for calculational purposes he can use wavefunction collapse, 
that is, the future evolution of his wavefunction is governed for the situation in which the initial setup for the spin was 
$(1 \ket{\uparrow} + 0 \ket{\downarrow})$.
The result in Eq.(\ref{FinalTwoPersonWavefunction}) correctly describes the experimental situation 
including the incorrect beliefs of the spectator 
and the feeling by both humans that the outcome is classical and unique 
both for the first term and for the second term in Eq.(\ref{FinalTwoPersonWavefunction}) (when this equation is expanded). 

When experimentalist A measures the spin of a spin-$1 \over 2$ object $N$ times, 
each time setting up the initial conditions in the same way so that the initial spin 
is always the $ (a_{\uparrow}  \ket{\uparrow} +  a_{\downarrow} \ket{\downarrow} )$ 
for fixed constants $a_{\uparrow}$ and $a_{\downarrow}$, 
it is natural to display the final wavefunction as a superposition of $2^N$ terms. 
For example, the wavefunction after the first measurement has the form 
$a_{\uparrow} \Psi'_{\uparrow} +  a_{\downarrow} \Psi'_{\downarrow}$. 
After the second measurement, it is 
$a_{\uparrow}(a_{\uparrow}\Psi''_{{\uparrow}{\uparrow}} + a_{\downarrow} \Psi''_{{\uparrow}{\downarrow}}) 
+ a_{\downarrow}(a_{\uparrow}  \Psi''_{{\downarrow}{\uparrow}} +a_{\downarrow} \Psi''_{{\downarrow}{\downarrow}})
$, 
where the first (respectively, second) term has evolved from $a_{\uparrow} \Psi'_{\uparrow}$ 
(respectively, $ a_{\downarrow} \Psi'_{\downarrow}$).
The final form after 3 measurements is
\begin{equation}
a_{\uparrow}^3\Psi'''_{{\uparrow}{\uparrow}{\uparrow}} + 
a_{\uparrow}^2 a_{\downarrow} (\Psi'''_{{\uparrow}{\uparrow}{\downarrow}} + \Psi'''_{{\uparrow}{\downarrow}{\uparrow}} + 
    \Psi'''_{{\downarrow}{\uparrow}{\uparrow}}) +
a_{\uparrow} a_{\downarrow}^2 (\Psi'''_{{\uparrow}{\downarrow}{\downarrow}} + \Psi'''_{{\downarrow}{\uparrow}{\downarrow}} + 
    \Psi'''_{{\downarrow}{\downarrow}{\uparrow}}) +
a_{\downarrow}^3\Psi'''_{{\downarrow}{\downarrow}{\downarrow}}
 \ ,
\label{MultipleSpinMeasurements} 
\end{equation}
and so on.
In the above, all the $\Psi$ are all functions of $\{ x_A \}$, $\{ x_o\}$ and $\{ x_e \}$. 
The wavefunction in Eq.(\ref{MultipleSpinMeasurements}) provides a correct description of the final situation: 
8 possible experimental spin outcomes each with the correct probability of happening. 
In each term, the experimentalist cannot be aware of the other 7 wavefunction components of the superposition 
(due to The Measurement Unawareness Result)
and he has a convincing feeling that the outcome is classical and unique. 
For example, for the term $a_{\uparrow}^2 a_{\downarrow}\Psi'''_{{\uparrow}{\uparrow}{\downarrow}}$ 
the experimentalist believes that he measured the spin to be up, up and down in that order.
If unitary quantum field theory is correct, 
then it is fair for the experimentalist to ask before the experiment has taken place, 
``Which of the above 8 terms will I be?''
The answer is that one does not know, which conforms to our expectations about and the uncertainty of quantum mechanics. 
However, when the experimentalist has completed the experiment, 
he knows what he observed due to The Empirical Consciousness Statement 
and therefore which of the above 8 terms he belongs to. 
So when he measures up, up and down, 
his future is based solely on the wavefunction associated with second term in Eq.(\ref{MultipleSpinMeasurements}), 
just as though the wavefunction had collapsed to the ${{\uparrow}{\uparrow}{\downarrow}}$ term. 

Unitary quantum field theory provides an unambiguous answer 
to the question of quantum immortality,\cite{QuantumImortalitySquires,QuantumImortalityMoravec,QuantumImortalityLewis,Tegmark}
something that a few people believe arises in the Many Worlds Interpretation of quantum mechanics. 
Consider the experiment in the previous paragraph modified so that if the spin is measured to be down 
then the experimentalist is killed by some means. 
If the experimentalist dies, then we also arrange for someone to remove the body from the scene of the experiment 
to a faraway location. 
The form of the wavefunction after the experimentalist has played one round of this ``quantum Russian roulette game'' is 
$a_{\uparrow} \Psi'_{\uparrow} (\{ x_A \}, \{ x_o\}, \{ x_e \}) +  a_{\downarrow} \Psi'_{\downarrow} (\{ x_A \}, \{ x_o\}, \{ x_e \})$. 
where $ \Psi'_{\uparrow}$ involves a wavefunction that corresponds to a live experimentalist, 
who knows he is alive and knows that the experiment measured an up spin, 
and $\Psi'_{\downarrow}$ involves a wavefunction that corresponds to a dead experimentalist. 
After the second round is conducted, the wavefunction becomes 
\begin{equation}
\begin{aligned}
a_{\uparrow}&
   (    a_{\uparrow}\Psi''_{{\uparrow}{\uparrow}} (\{ x_A \}, \{ x_o\}, \{ x_e \}) + 
   a_{\downarrow} \Psi''_{{\uparrow}{\downarrow}} (\{ x_A \}, \{ x_o\}, \{ x_e \}) )   \\
+ a_{\downarrow}& \Psi''_A (\{ x_A \}, \{ x_o\} ) 
    (a_{\uparrow}  \Psi''_{{\downarrow}{\uparrow}} (\{ x_o\}, \{ x_e \}) + 
  a_{\downarrow} \Psi''_{{\downarrow}{\downarrow}} (\{ x_o\}, \{ x_e \}) )
 \ .
\label{SecondRound} 
\end{aligned}
\end{equation}
Note that the wavefunction $ \Psi''_A (\{ x_A \}, \{ x_o\} )$ for the (dead) experimentalist 
in the second term of Eq.(\ref{SecondRound}) 
is common to both the  terms $\Psi''_{{\downarrow}{\uparrow}}$ and $\Psi''_{{\downarrow}{\downarrow}}$
because the experimentalist (who died in the first round) 
cannot be affected by the outcome of the second round of the experiment. 
The wavefunction $\Psi''_{{\uparrow}{\uparrow}}$ involves a live experimentalist, 
who knows he is alive and knows that the experiment measured an up spin twice, 
while  $\Psi''_{{\uparrow}{\downarrow}}$ involves a wavefunction that corresponds to an experimentalist 
who survived the first round but died in the second round. 
It straightforward to determine the structure of the wavefunction after $N$ rounds of quantum Russian roulette 
and the probability that an experimentalist survives is $| a_{\uparrow} |^{2N}$. 
The idea of quantum immortality arises in the Many Worlds Interpretation of quantum mechanics 
if one attaches a ``world'' to each of the terms 
in the linear superposition and considers the worlds to be simultaneously present, 
meaning that in one of the worlds (the one corresponding to $N$ up spin measurements, namely, $\Psi^{(N)}_{{\uparrow} \dots {\uparrow}}$) 
the experimentalist is still alive. 
However, in unitary quantum field theory, the terms in the wavefunction represent a logical `OR' condition 
so that the worlds are not simultaneously present: 
either the experimentalist is alive and knows he survived and measured all the spins to be up, 
which corresponds to the term proportional to $\Psi^{(N)}_{{\uparrow} \dots {\uparrow}}$  in the superposition, 
{\it or} he died in one of the $N$ rounds. 

We now address in unitary quantum field theory the issue of 
constructive and destructive quantum mechanical interference in experiments. 
Consider a screen, such as the one that is used in the two-split experiment, 
that can detect the position $\vec x_O$ of a quantum object $O$ by a burst of photons. 
For simplicity, consider two very small localized regions on the screen $A$ and $B$, 
or, alternatively, use a screen in which only these two detection regions are present. 
Consider a wavefunction of the form $\phi_1(\vec x_O, t) = ( \phi_A(\vec x_O, t) +  \phi_B(\vec x_O, t))/\sqrt{2}$ 
where, at times $t$ when the wavefunction is near the detection screen
(but has not yet interacted with it), 
the support of $\phi_A (\vec x_O, t)$  (respectively, $\phi_B(\vec x_O, t)$) is near the region $A$ (respectively, $B$):
$\phi_A (\vec x_O, t) = 0$ when $\vec x_O$ is near region $B$ and $\phi_B(\vec x_O, t) = 0$ when $\vec x_O$ is near region $A$.
When experimentalist $H$ conducts the ``detection'' experiment for the quantum mechanical object 
for a setup in which the relevant initial wavefunction is $\phi_1$, 
the final form of the wavefunction is the generalized Schr\"odinger Cat superposition 
\begin{equation}
\phi'_1 =  {{ \Psi'_A (\{ x_H \}, \vec x_O, \{ x_o\},  \{ x_e\}, t) +  \Psi'_B (\{ x_H \}, \vec x_O, \{ x_o\},  \{ x_e\}, t) } \over {\sqrt{2}}}
 \ , 
\label{FinalWavefunction1} 
\end{equation}
where $\{ x_H \}$ are the degrees of freedom of the experimentalist, 
and $\Psi'_A$ (respectively, $\Psi'_B$) involves a wavefunction 
in which the experimentalist has seen light originating from region $A$ (respectively, $B$).
Now consider repeating the experiment in which, 
due to either different initial conditions or a different initial setup,  
the relevant initial wavefunction for $O$ is 
$\phi_2 (\vec x_O, t) = ( \phi_A(\vec x_O, t) - \phi_B(\vec x_O, t))/\sqrt{2}$. 
Then the final form of the wavefunction for this case is 
\begin{equation}
\phi'_2 =  { { \Psi'_A (\{ x_H \}, \vec x_O, \{ x_o\},  \{ x_e\}, t) - \Psi'_B (\{ x_H \}, \vec x_O, \{ x_o\},  \{ x_e\}, t) } \over {\sqrt{2}}}
 \ . 
\label{FinalWavefunction1} 
\end{equation}
Finally consider the situation in which the initial wavefunction is $\phi = (\phi_1 + \phi_2) /\sqrt{2}$. 
At times when the wavefunction is near the detection screen, 
there is constructive quantum interference near the region $A$ and destructive interference near the region $B$. 
The spatial probability distribution is 
$| \phi (\vec x)|^2 = (| \phi_1 (\vec x)|^2 +  | \phi_2 (\vec x)|^2 + \phi_1^* (\vec x) \phi_2 (\vec x) + \phi_1 (\vec x) \phi_2^* (\vec x) )/2$. 
The ``interference'' term $(\phi_1^* (\vec x) \phi_2 (\vec x) + \phi_1 (\vec x) \phi_2^* (\vec x) )/2$ 
is exactly what is needed to cancel out the support of $\phi$ in the region near $B$.
In the quantum mechanical double-slit experiment, 
$\phi_1$ might be associated with one slit and $\phi_2$ with the other slit,  
and region $A$ (respectively, $B$) might be associated with a place on the screen with constructive (respectively, destructive) interference.
Now when one looks at the situation at a later time when $\phi' = (\phi'_1 + \phi'_2) /\sqrt{2}$, 
one sees the same constructive and destructive interference:
the effect of the interference term 
$(\phi_1^{'*}  (\{ x_H \}, \vec x_O, \{ x_o\},  \{ x_e\}, t) \phi'_2 (\{ x_H \}, \vec x_O, \{ x_o\},  \{ x_e\}, t) + 
\phi'_1 (\{ x_H \}, \vec x_O, \{ x_o\},  \{ x_e\}, t) \phi_2^{'*}  (\{ x_H \}, \vec x_O, \{ x_o\},  \{ x_e\}, t) )/2$ 
is to remove the possibility of the experimentalist seeing the light coming from region $B$. 
Note that this involves the quantum mechanical interference of two wavefunctions 
each of which is associated with a generalized Schr\"odinger Cat.
The interference effects at the ``microscopic'' level (that is, in terms of wavefunction $\phi$ of the object $O$)
are transmitted to the ``macroscopic'' level 
(that is, in terms of $\phi'$, 
which involves the degrees of freedom of the experimentalist $\{ x_H \}$, the object $ \vec x_O$, 
the experimental device $ \{ x_e\}$ and any other $ \{ x_o\}$ relevant degrees of freedom). 
The above statements apply to the situation in which there is ``partial'' interference 
(obtained from the above by having the magnitude of the coefficients 
of $\phi_A$ and $\phi_B$ in the expansions of $\phi_1$ and $\phi_2$ be different from $1 \over \sqrt{2}$). 
These results all follow from linearity.

\end{document}